\documentclass[aps,prb,twocolumn,floatfix]{revtex4}
\input epsf
\usepackage{graphicx}
\usepackage{bbm}

\def\tens#1{\underline{\underline{#1}}}
\begin{document}

\title{Ab-initio study of the vibrational properties of crystalline TeO$_2$: the $\alpha$, $\beta$,
and $\gamma$ phases}
\author{M. Ceriotti, F. Pietrucci,  and M. Bernasconi}

\affiliation{Dipartimento di Scienza dei Materiali 
 Universit\`a di  Milano-Bicocca, Via Cozzi 53, I-20125, Milano, Italy}

\begin{abstract}
Based on density functional perturbation theory, we have studied the vibrational properties of three
crystalline phases of tellurium dioxide: paratellurite $\alpha$-TeO$_2$, tellurite $\beta$-TeO$_2$ and
the  new phase  $\gamma$-TeO$_2$, recently identified experimentally.
Calculated Raman and IR spectra are in good agreement with available experimental data.
The vibrational spectra of $\alpha$- and $\beta$-TeO$_2$ can be interpreted in terms of vibrations of 
TeO$_2$ molecular units.
\end{abstract}

\maketitle

\section{Introduction}

\begin{figure}
\caption{The $\mathrm{TeO_4}$ bipyramidal unit, building block of the three $\mathrm{TeO_2}$ polymorphs.
The two small circles indicate the Te lone-pair. Dark (light) spheres indicate Te (O) atoms.}
\includegraphics[width=0.5\columnwidth]{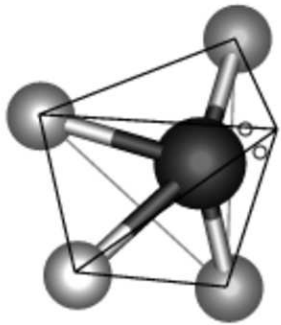}
\label{bipiramide}
\end{figure}

Tellurium oxide (TeO$_2$) and  TeO$_2$-based glasses are promising active materials  for optical switching 
devices due to their large non-linear polarizability\cite{tghandbook} and for optical amplifiers due to their large 
cross section for Stimulated Raman Scattering  (SRS).\cite{daitassone,ramangain}
Three crystalline phases of TeO$_2$ are well documented.
The paratellurite $\alpha$-TeO$_2$ \cite{thomas} and the tellurite $\beta$-TeO$_2$ \cite{beyer} phases have been known
for a long time. Recently, a third crystalline polymorph metastable at normal conditions, $\gamma$-TeO$_2$,
\cite{gamma1,gamma2}  has been identified by x-ray powder diffraction of recrystallized tellurite glasses doped with metal
oxides. The three phases can be described as different arrangements of corner sharing TeO$_4$
unit (the distorted trigonal bipyramids in Fig. \ref{bipiramide}). 
These units have two longer $\mathrm{Te-O}$ bonds (axial bonds)
and two shorter ones (equatorial bonds) whose length varies in the different phases.
Paratellurite, $\alpha$-TeO$_2$, has a 3-dimensional fully connected structure, while $\beta$-TeO$_2$
has a layered structure with weakly bonded sheets.
In $\gamma$-TeO$_2$,  one $\mathrm{Te-O}$ bond is substantially larger than the other three and  by breaking this latter
bond in a 3-dimensional visualization of the network, a chain-like structure appears.
The $\gamma$-TeO$_2$ phase has then been described in terms of a polymeric form made of TeO$_3$ units.\cite{gamma1,gamma2}
By comparing the Raman spectra of TeO$_2$ glasses with those of the crystalline phases, it has also
been proposed that the chain-like structure of $\gamma$-TeO$_2$ would represent the main structural
feature of the glass.\cite{fononi_beta}
Theoretical lattice dynamics calculations of the phonon spectra of crystalline TeO$_2$ 
have already appeared in literature, \cite{gamma2,fononi_gamma,fononi_beta,alfashell,tesi}
 based on empirical interatomic force constants from
two- and three body interactions or shell models. \cite{alfashell}
The lattice dynamics calculations revealed 
that the phononic spectra of $\alpha$-TeO$_2$ and $\beta$-TeO$_2$
can be well interpreted in terms of vibrations of weakly coupled TeO$_2$ molecules.\cite{gamma2,tesi,fononi_gamma,fononi_beta} Accordingly, tellurite and
paratellurite would be seen as molecular crystals made of TeO$_2$ molecules. The polymeric $\gamma$-TeO$_2$ would represent
a chain-like polymerization of TeO$_2$ molecules giving rise to a structure more connected than paratellurite and
tellurite phases.

In this paper, we  investigate further 
the three crystalline phases of TeO$_2$ ($\alpha$, $\beta$, $\gamma$) by computing their
  structural and vibrational properties from first principles.
We aim at providing a compelling assignment of the 
experimental Raman and IR peaks to specific phonons which would allow to identify the
vibrational signature of the different structural units in the crystals and in the glass.

After a brief description  of our theoretical framework in section II, we present our results on the structural
and electronic properties in section III and on the vibrational properties in section IV. Section V reports our
 conclusions.

\section{Computational details}

Calculations are performed within the framework of Density Functional Theory (DFT)  within 
the simple local density approximation (LDA) and with
 gradient corrected  exchange and correlation energy functionals  (PBE \cite{pbe} and BLYP \cite{blyp}),
as implemented in the codes PWSCF and PHONONS. \cite{pwscf}
Calculations are performed with ultrasoft pseudopotentials.\cite{vanderbilt}
Kohn-Sham (KS) orbitals are expanded in a plane waves basis up to a kinetic cutoff of 30  Ry.
Brillouin Zone (BZ) integration has been performed over  Monkhorst-Pack (MP) meshes of
 8$\times$8$\times$8, 4$\times$4$\times$4 and 6$\times$6$\times$6  for $\alpha$-, $\beta$-, 
and $\gamma$-TeO$_2$, respectively. \cite{MP}
Equilibrium geometries have been obtained by optimizing the internal and lattice structural parameters at 
several volumes and fitting the energy versus volume data with a Murnaghan function. \cite{murna}
Residual anisotropy in the stress tensor at the optimized lattice parameters at each volume is below 0.6 kbar.
Infrared and Raman spectra are obtained from effective charges, dielectric susceptibilities and 
phonons at the $\Gamma$ point within density functional perturbation theory. \cite{dfpt}
Effective charges have been computed with norm-conserving pseudopotentials,\cite{TM} 
and a kinetic cutoff of 70 Ry.
Relevant formula for the calculation of the IR and Raman spectra
are given in section IV.
Preliminary tests on the transferability of the pseudopotentials and on the accuracy of the exchange and 
correlation functionals have been performed on the structure and vibrational frequencies of the TeO$_2$ molecule. 
Results are presented in Table \ref{molecola} for PBE and BLYP functionals and compared with experimental 
data and previous B3LYP all-electrons calculations.

\begin{table}
\caption{Bond length, bond angle and vibrational frequencies  of $\mathrm{TeO_2}$ molecule, compared with previous 
 B3LYP\protect\cite{alfa_ae} and experimental data \protect\cite{exp_mol}.  $B_1$ and $A_1$(2) are symmetric and antisymmetric
stretching modes, respectively, while $A_1$(1) is a bending mode of the TeO$_2$ molecule.}
\label{molecola}
\begin{ruledtabular}
\begin{tabular}[c]{l c c c c c}
 & &  & \multicolumn{3}{c}{$\omega (cm^{-1})$}\\
                   & Te-O (\AA) & Te-O-Te (deg.) & $A_1(1)$ & $A_1(2)$ &  $B_1  $ \\\hline
PBE                & 1.832      & 111.7          & 266      & 804      & 806      \\
BLYP               & 1.820      & 111.8          & 265      & 813      & 824      \\
B3LYP              & 1.844      & 112.9          & 250      & 883      & 921      \\
Exp                & 1.840      & 112            & 294      & 810      & 849      \\
\end{tabular}
\end{ruledtabular}
\end{table}

\section{Structural properties}

\subsection{$\alpha$-TeO$_2$}

\begin{figure}
\caption{Structure of $\alpha$-$\mathrm{TeO}_2$. (a) Perspective view, two unit cells are shown;
O2 and O3 (O1 and O4) label the equatorial (axial) oxygen atoms. (b)  Four unit cells are drawn along $c$ axis.}
\includegraphics[width=1.0\columnwidth]{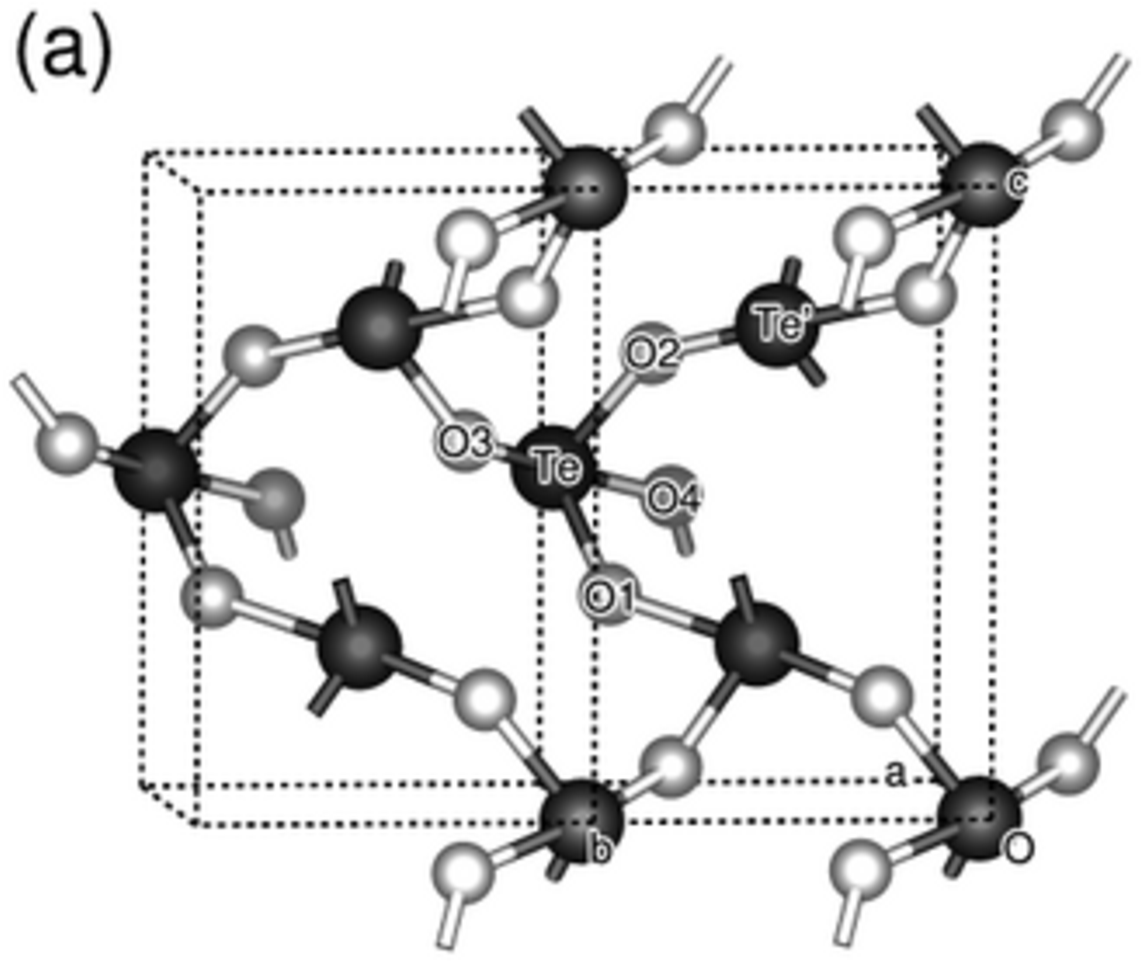}
\includegraphics[width=1.0\columnwidth]{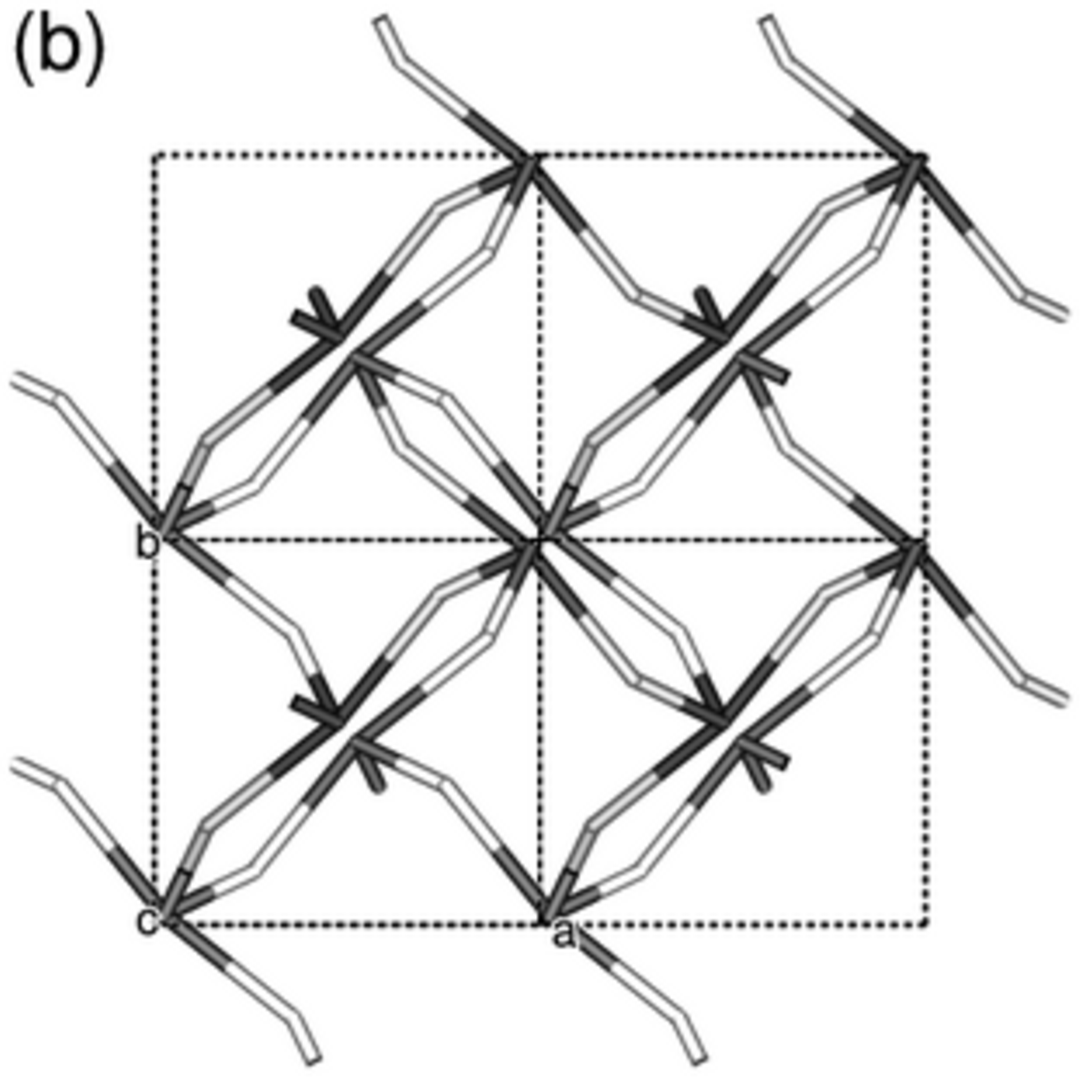}
\label{strut_alfa}
\end{figure}

The paratellurite crystal (space group $P4_12_12$, $D_4^4$, n. 92) has tetragonal symmetry with four
formula units per unit cell. \cite{thomas} Two atoms are independent by symmetry: one Te atom at   position $(x,x,0)$
and one oxygen atom at position $(\alpha,\beta,\gamma)$. The experimental and theoretical
(LDA and PBE) lattice parameters and position of the symmetry independent atoms are given in table \ref{pos_alfa}.
As usual the LDA (GGA) underestimates (overestimates) the equilibrium volume with respect to experimental data.
The bulk modulus and its derivative  with respect to pressure obtained from the fitted Murnaghan
equation of state (PBE functional) are $B$=33.0 GPa and $B'$=8.1 to be compared with 
the experimental value of  $B$=45.3 GPa. \cite{bmodulus}
These results are in good agreement with previous all-electron calculations which confirms
the transferability of our pseudopotentials. \cite{kleinman}
Concerning the internal structure of $\alpha$-TeO$_2$, a Te atom is coordinated with four oxygen atoms,
two equatorial atoms at closer distance (exp. 1.88 \AA) and two axial atoms at larger distance
(exp. 2.12 \AA).\cite{thomas} The structure can be seen as a network of corner-sharing TeO$_4$ units
(Fig. \ref{strut_alfa}), the oxygen atoms bridging two Te atoms. The network is formed by
six-membered rings of Te atoms with bridging oxygen atoms.
As already mentioned, lattice dynamics calculations with empirical interatomic potentials
suggest an alternative picture of $\alpha$-TeO$_2$ as formed by weakly interacting TeO$_2$ molecules. \cite{tesi,fononi_gamma,fononi_beta}
The calculated independent bond lengths and angles are compared with experimental data in Table \ref{bond_alfa}
for LDA and PBE functionals at the theoretical equilibrium lattice parameters, and for PBE and BLYP
functionals at the experimental lattice parameters.
Although the error in the lattice parameters is within the expected accuracy for DFT calculations, the 
error in the bond lengths at the theoretical lattice parameters is somehow larger than usual (cfr. Table 
\ref{bond_alfa}).

\begin{table}
\caption{Structural parameters of $\alpha\mathrm{-TeO_2}$, calculated  with  
 LDA functional, with  PBE functional at the theoretical (PBE) and experimental 
(PBEexp) lattice parameters, and  with BLYP functional at the experimental 
 lattice parameters (BLYPexp). Experimental X-ray data (Exp) are from Ref. \protect\cite{thomas}.}
\label{pos_alfa}
\begin{ruledtabular}
\begin{tabular}[c]{l c c c c c}
                     & LDA       & PBE     & PBEexp& BLYPexp &  Exp   \\\hline
\multicolumn{6}{l}{Cell parameters (\AA)}\\\hline
a                    & 4.819     & 4.990  &   -     &    -      & 4.8082   \\
c                    & 7.338     & 7.546   &   -     &    -      & 7.612    \\\hline
\multicolumn{6}{l}{Atomic positions}\\\hline
Te $x$               & 0.0025    & 0.0272  & 0.0156  & 0.0201    & 0.0268   \\
O  $\alpha$          & 0.1402    & 0.1467  & 0.1371  & 0.1388    & 0.1368   \\
O  $\beta$           & 0.2379    & 0.2482  & 0.2492  & 0.2519    & 0.2576   \\
O  $\gamma$\hspace{1cm} & 0.2065 & 0.1968  & 0.1954  & 0.1926    & 0.1862   \\
\end{tabular}
\end{ruledtabular}
\end{table}

\begin{figure}
\caption{(a) Lattice parameters ($a$, $c$), 
(b) bond lengths and (c) bond angles of $\alpha$-$\mathrm{TeO_2}$ as a function of unit cell
volume. The arrow indicates the theoretical equilibrium value.
The series in panel (a) correspond (from bottom to top) to $a$ and $c$ axes. 
$\mathrm{Te-O3}$ and $\mathrm{Te-O1}$ are the bond lengths of the axial and equatorial bonds, respectively.}
\includegraphics[width=1.0\columnwidth]{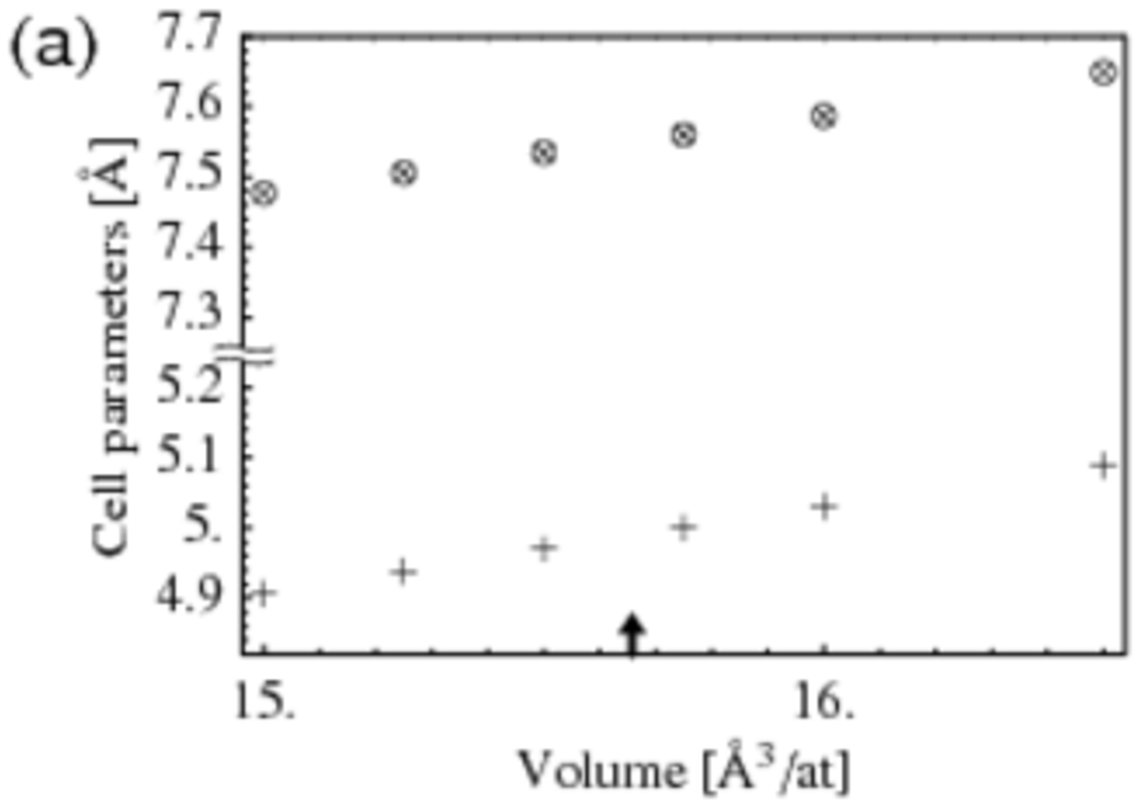}
\includegraphics[width=1.0\columnwidth]{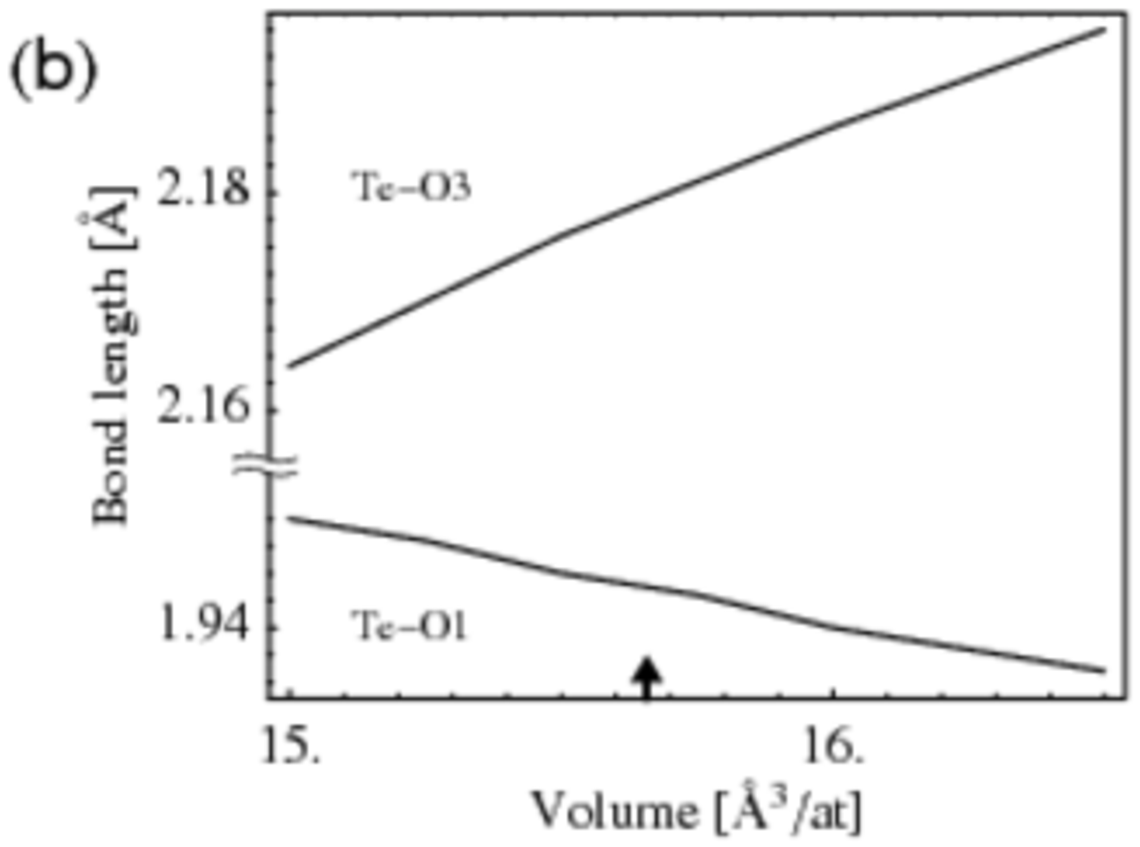}
\includegraphics[width=1.0\columnwidth]{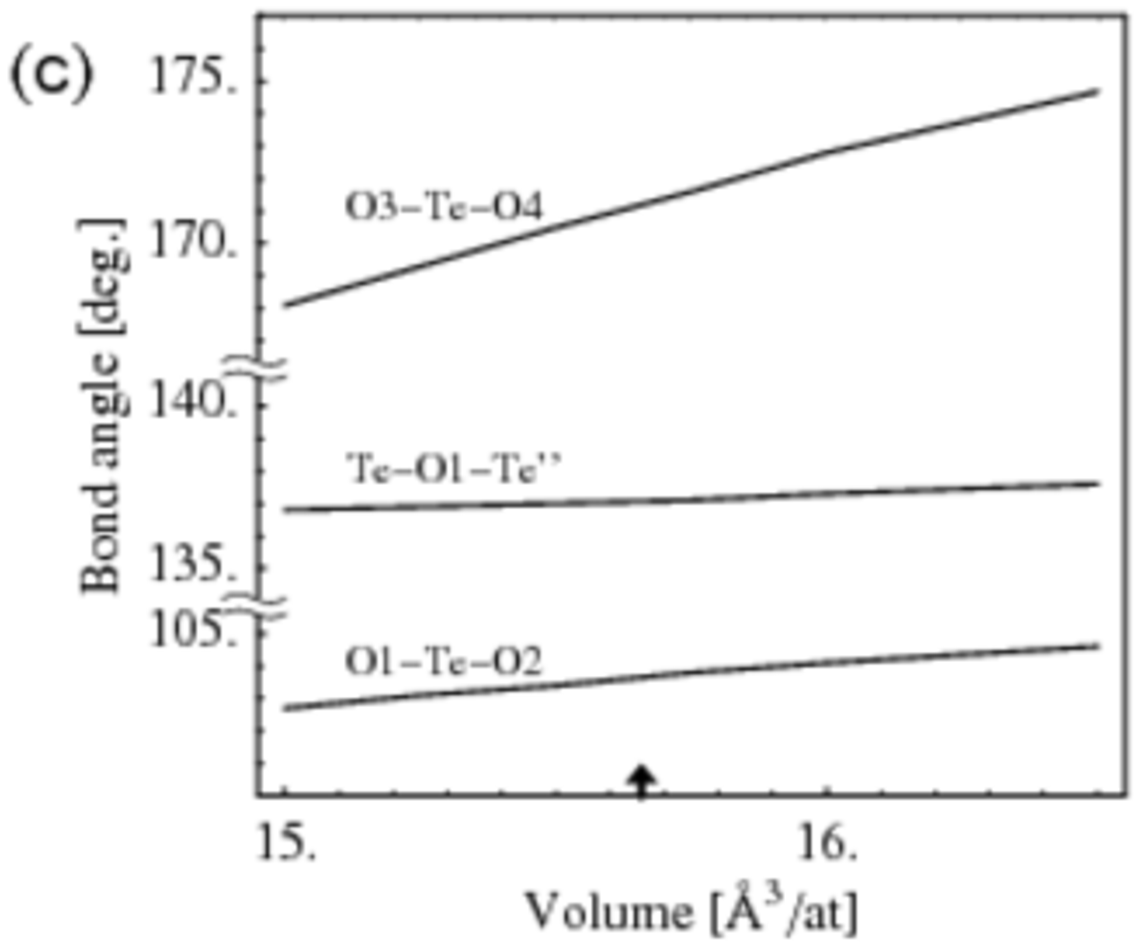}
\label{assi_alfa}
\end{figure}

In particular the calculated equatorial and axial $\mathrm{Te-O}$ bond lengths are much closer to each other than in the
experimental structure. This  misfit is larger for LDA than for PBE functional, but it is sizably reduced by
optimizing the internal geometry at the experimental lattice parameters with a marginal improvement for
the BLYP functional (cfr. Table \ref{bond_alfa}). 
The change in the lattice parameters, bond lengths and bond angles with volume is reported in Fig. \ref{assi_alfa}.
The increase of the $c/a$ ratio with pressure is in agreement with the experimental data of Ref. \cite{pressure_alfa}.
By increasing the volume, the length of the shorter equatorial bonds decreases while the longer axial bonds increases in length.
This behavior would support the picture 
of $\alpha$-TeO$_2$ 
as made of TeO$_2$ molecules, 
inferred from 
previous  lattice dynamics calculations \cite{tesi,fononi_gamma,fononi_beta} which turns out to be in good agreement
with our ab-initio phonon calculations reported in Sec. IV.
Although useful for the analysis of the vibrational spectra, the picture of paratellurite as a molecular crystal 
does not account for all the properties of $\alpha$-TeO$_2$. For instance, it can be hardly conciliated with the mostly ionic character 
and electronic bandwidths of several eV emerged from previous ab-initio calculations \cite{kleinman,palmero}.

\begin{table}
\caption{Bond lengths and angles for $\alpha\mathrm{-TeO_2}$, 
calculated  with              
 LDA functional, with  PBE functional at the theoretical (PBE) and experimental 
(PBEexp) lattice parameters, and  with BLYP functional at the experimental     
 lattice parameters (BLYPexp). Experimental X-ray data (Exp) are from Ref. \protect\cite{thomas}.
  Atoms are labelled according to Fig. \protect\ref{strut_alfa}.}
\label{bond_alfa}
\begin{ruledtabular}
\begin{tabular}[c]{l c c c c c}
                     & LDA       & PBE     & PBEexp& BLYPexp &  Exp   \\\hline
\multicolumn{6}{l}{Bond lengths (\AA)}\\\hline
Te-O1                & 2.028     & 1.944   & 1.954   & 1.924     & 1.878    \\
Te-O3                & 2.170     & 2.118   & 2.145   & 2.134     & 2.122    \\
Te-Te'\hspace{1cm}   & 3.855     & 3.838   & 3.806   & 3.779     & 3.742    \\\hline
\multicolumn{6}{l}{Angles (degrees)}\\\hline
O1-Te-O2             &   99.8    &  103.6  &  103.6  &  103.7    &  103.4   \\
O3-Te-O4             &  161.5    &  171.2  &  165.3  &  166.7    &  168.0   \\
Te-O2-Te'            &  135.0    &  137.1  &  136.4  &  137.2    &  138.6   \\
\end{tabular}
\end{ruledtabular}
\end{table}

\subsection{$\beta$-TeO$_2$}

\begin{figure}
\caption{Layered structure of $\beta$-$\mathrm{TeO_2}$. (a) 
A projection along $ab$ plane. (b)  Sketch 
 of a single layer. O1 and O2 ($\mathrm{O1'}$ and $\mathrm{O2'}$) label equatorial (axial) atoms.}
\includegraphics[width=1.0\columnwidth]{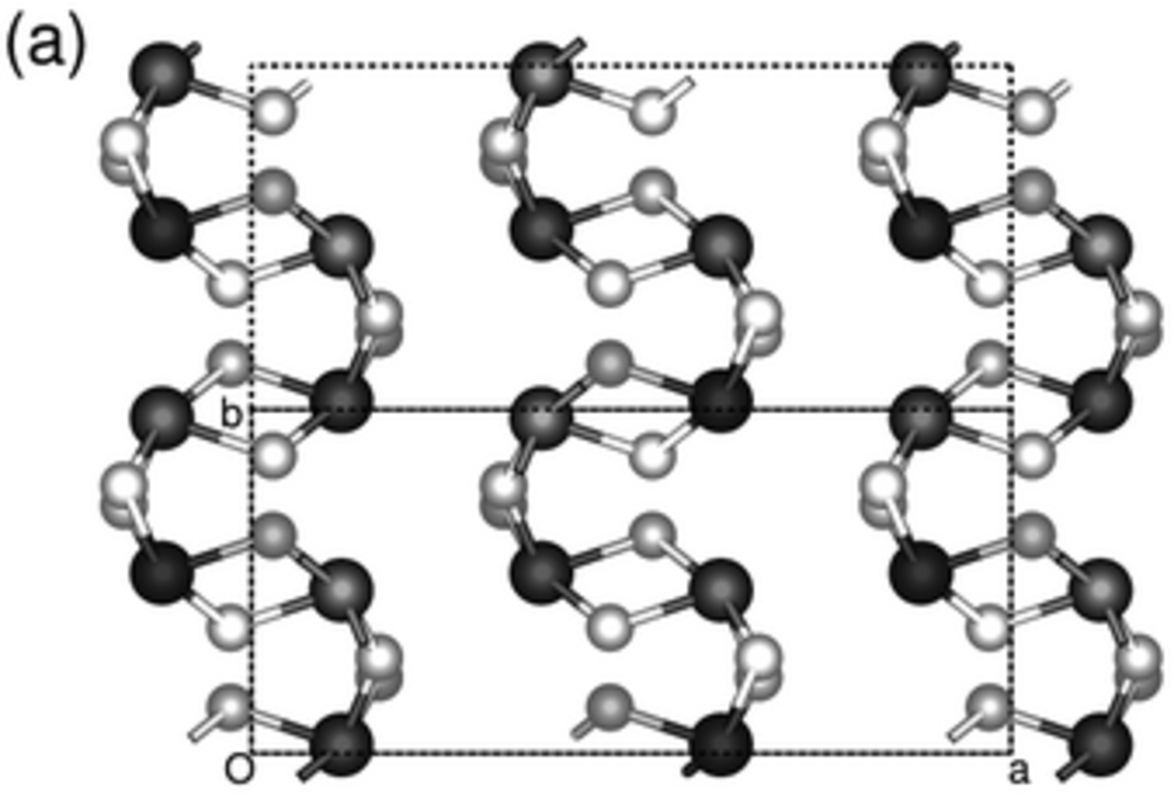}
\includegraphics[width=1.0\columnwidth]{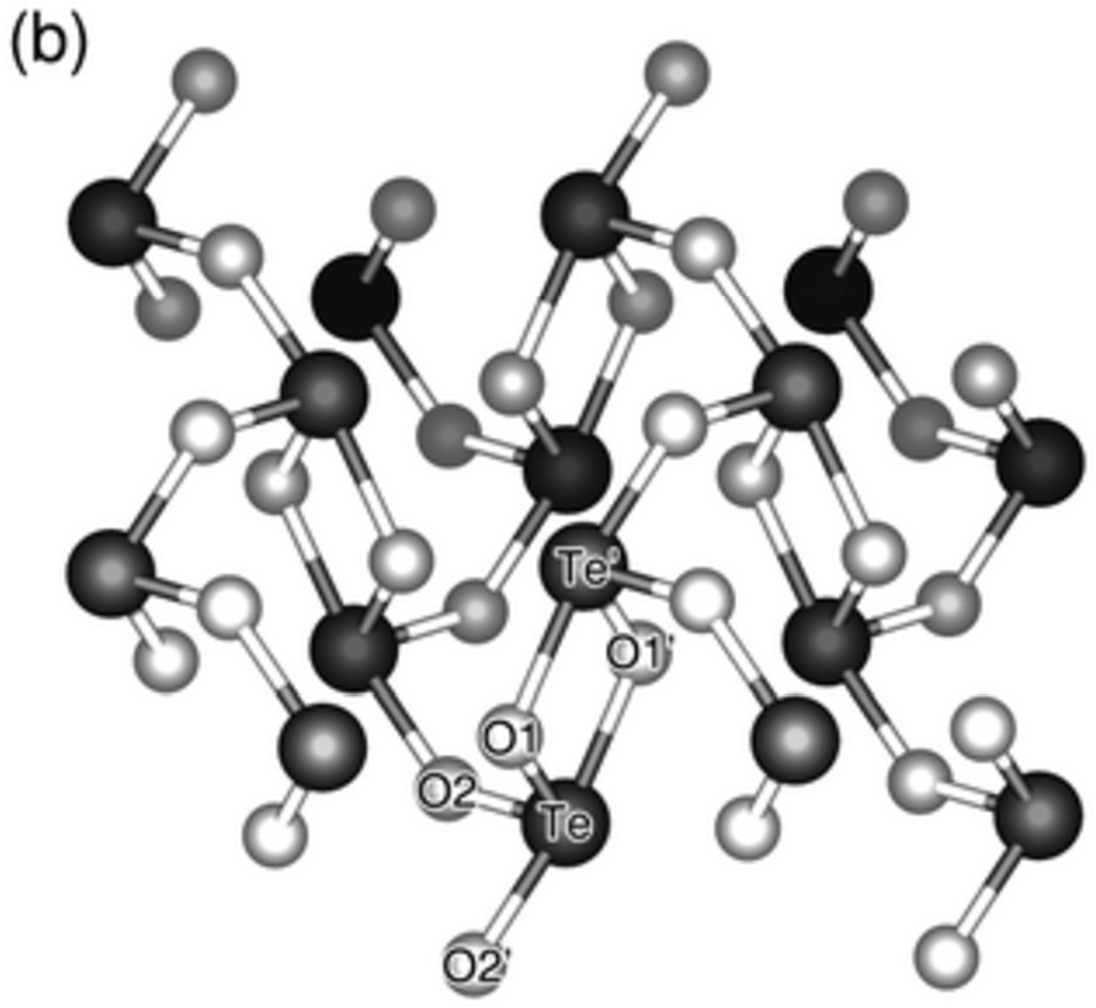}
\label{strut_beta}
\end{figure}

The tellurite crystal (space group $Pbca$, $D_{2h}^{15}$, n. 61) has orthorhombic symmetry with eight
formula units per unit cell.\cite{beyer} Two oxygen and one tellurium atoms are independent by symmetry.
The experimental and theoretical (PBE) lattice parameters and positions of the symmetry independent atoms 
are given in table \ref{pos_beta}.
The structural unit of $\beta$-TeO$_2$ is still a TeO$_4$ trigonal bipyramid, as in $\alpha$-TeO$_2$, which forms a 
network by sharing vertices (bridging oxygen atoms). However, as opposed to paratellurite, the $\beta$-TeO$_2$ 
crystal has a layered structure with layers oriented perpendicular to the $a$ axis (Fig. \ref{strut_beta}) 
and an interlayer distance of $\sim$ 3 \AA. \cite{beyer}
We can still recognize two longer axial bonds and two shorter equatorial bonds, but here all the four $\mathrm{Te-O}$ bonds of the
TeO$_4$ unit differ in length.
The 2D  network is formed by two- and six-membered Te rings. Still, $\beta$-TeO$_2$ as well can be seen as a molecular crystal
made of TeO$_2$ molecules according to lattice dynamics calculations. \cite{tesi,fononi_gamma,fononi_beta}

\begin{table}
\caption{Structural parameters of $\beta\mathrm{-TeO_2}$, computed 
using PBE functional at the theoretical (PBE) and experimental
(PBEexp) lattice parameters, and  with BLYP functional at the experimental
 lattice parameters (BLYPexp). Experimental X-ray data  (Exp) are from Ref.
 \protect\cite{beyer}.}
\label{pos_beta}
\begin{ruledtabular}
\begin{tabular}[c]{l c c c c}
                      & PBE     & PBEexp& BLYPexp &  Exp   \\\hline
\multicolumn{5}{l}{Cell parameters (\AA)}\\\hline
a                     & 12.232  &    -    &     -     & 12.035   \\
b                     & 5.523   &    -    &     -     & 5.464    \\
c                     & 5.803   &    -    &     -     & 5.607    \\\hline
\multicolumn{5}{l}{Atomic positions}\\\hline
Te $x$                & 0.1173  & 0.1193  & 0.1189    & 0.1181   \\
Te $y$                & 0.0263  & 0.0184  & 0.0211    & 0.0252   \\
Te $z$                & 0.3748  & 0.3677  & 0.3682    & 0.3378   \\
O(1) $x$              & 0.0314  & 0.3235  & 0.0318    & 0.028    \\
O(1) $y$              & 0.6491  & 0.6422  & 0.6412    & 0.634    \\
O(1) $z$              & 0.1613  & 0.1721  & 0.1703    & 0.171    \\
O(2) $x$              & 0.1714  & 0.1743  & 0.1723    & 0.168    \\
O(2) $y$              & 0.2174  & 0.2108  & 0.2124    & 0.221    \\
O(2) $z$ \hspace{1cm} & 0.0812  & 0.0694  & 0.0714    & 0.081    \\
\end{tabular}
\end{ruledtabular}
\end{table}

\begin{figure}
\caption{(a) Lattice parameters ($a$, $b$, $c$), (b) bond lengths and (c) bond angles of $\beta$-$\mathrm{TeO_2}$ as a function of unit cell
volume. The arrow indicates the theoretical equilibrium value. 
 The three series in panel (a) correspond (from bottom to top) to $b$, $c$ and $a$ axes.
The atom labels refer to Fig. \protect\ref{strut_beta}.
$\mathrm{Te-O1}$ and $\mathrm{Te-O2}$ are equatorial bonds, while $\mathrm{Te-O1'}$ and $\mathrm{Te-O2'}$ are axial bonds.}
\includegraphics[width=1.0\columnwidth]{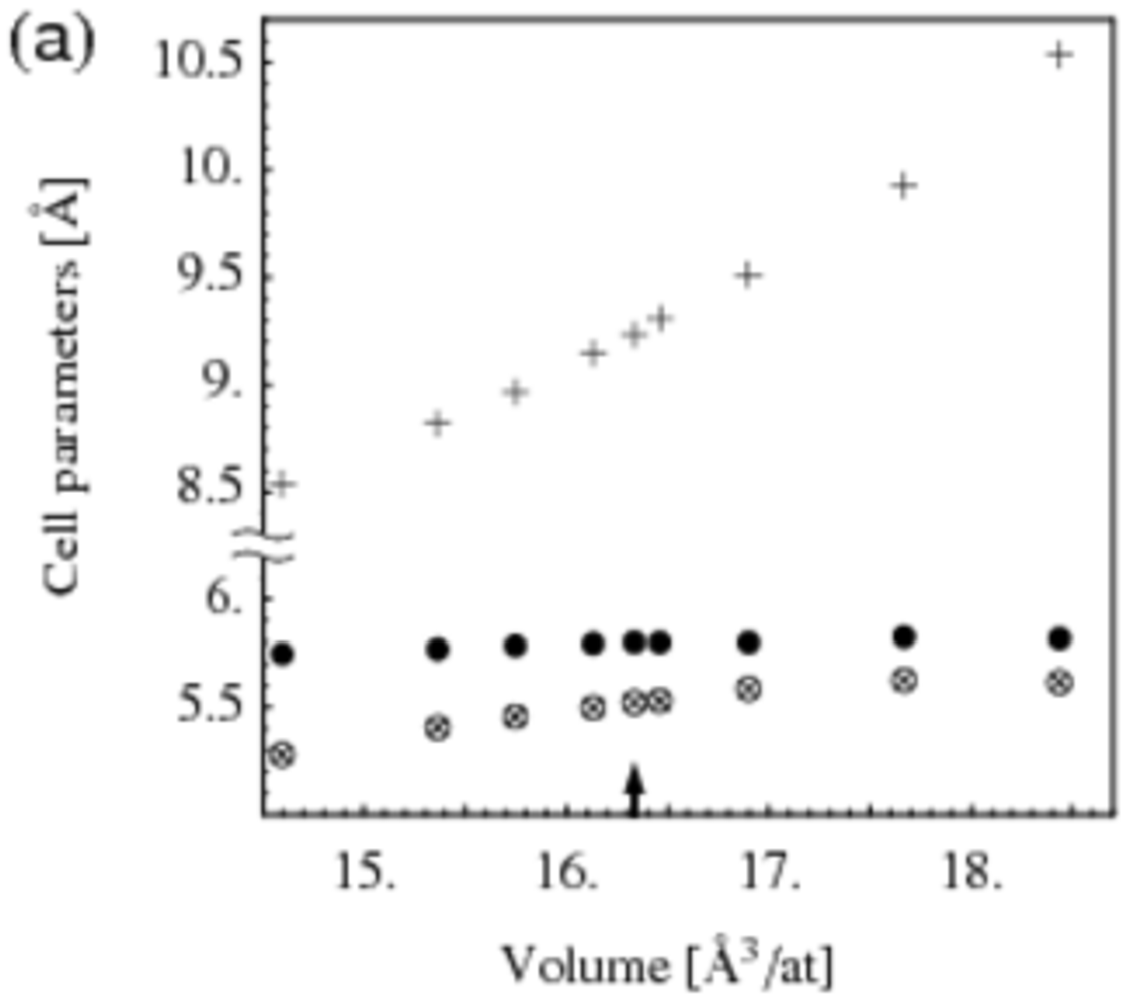}
\includegraphics[width=1.0\columnwidth]{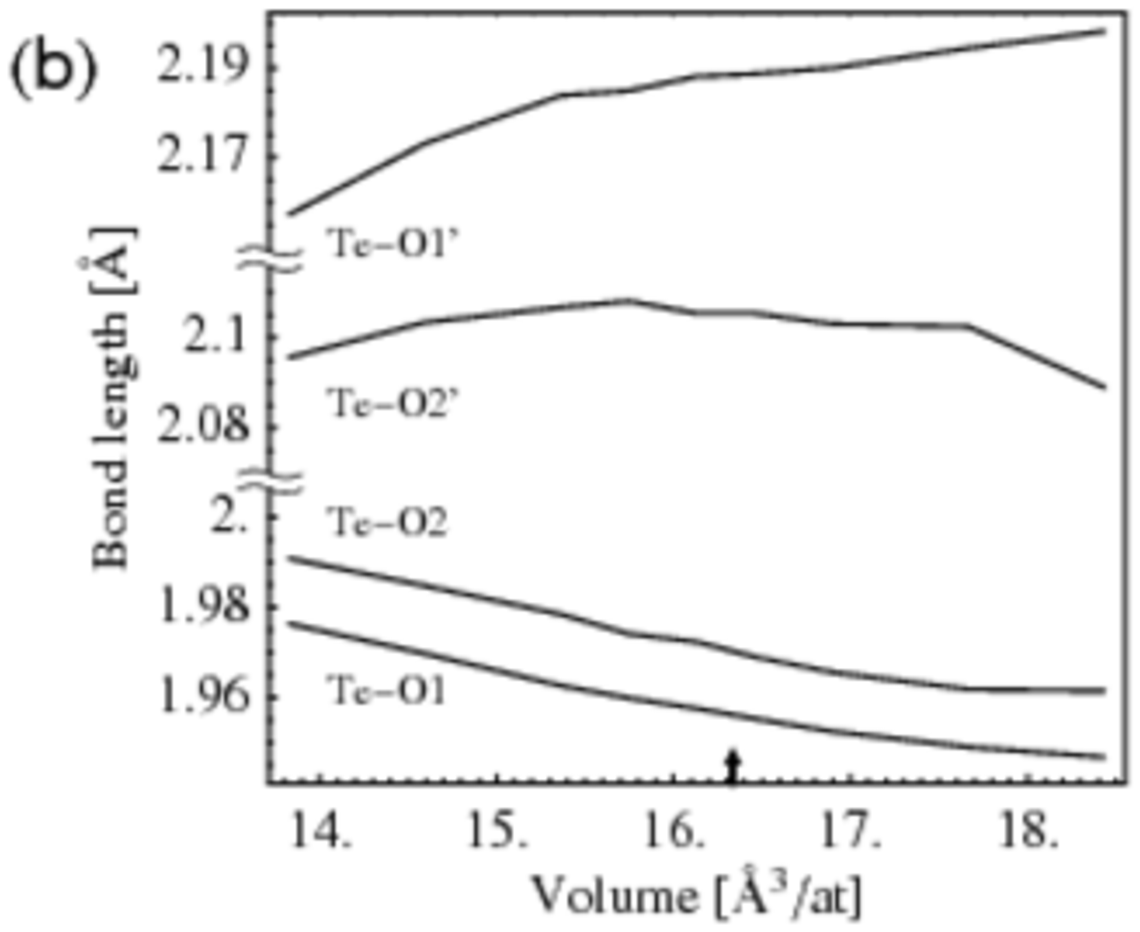}
\includegraphics[width=1.0\columnwidth]{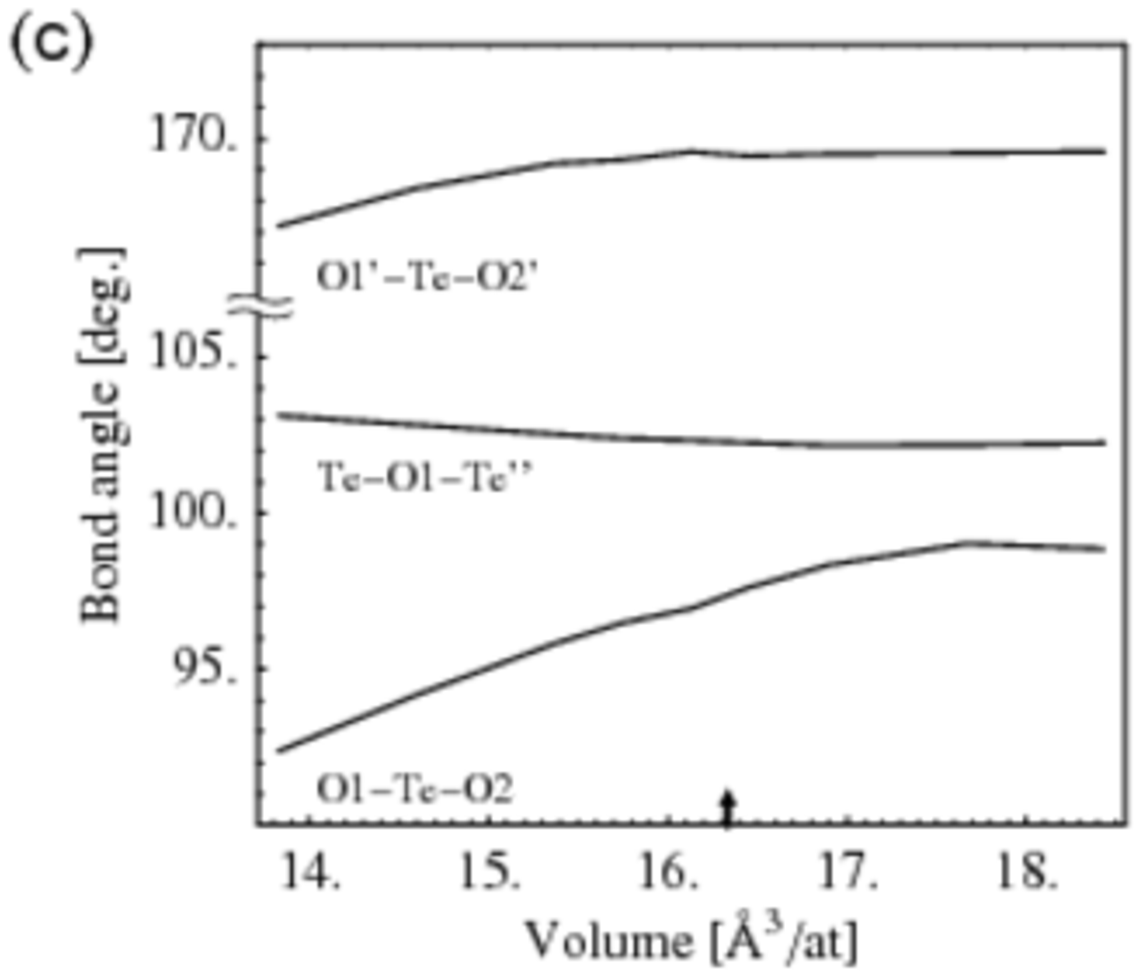}
\label{varvol_beta}
\end{figure}

The calculated independent bond lengths and angles are compared with experimental data in Table \ref{bond_beta}
for  PBE functional at the theoretical equilibrium lattice parameters, and for PBE and BLYP
functionals at the experimental lattice parameters.
The same comments on the internal structure presented for $\alpha$-TeO$_2$ holds here for $\beta$-TeO$_2$.
The spread in the bond lengths  of the theoretical geometry  at the theoretical lattice parameters (PBE)
is reduced with respect to experiments. The misfit is reduced by optimizing the internal geometry at the 
experimental lattice parameters, with a marginal improvement for the BLYP  over PBE functional.
The error in the internal structure thus comes primarily from the overestimation of the equilibrium volume 
(6 $\%$) by PBE functional. 
A possible source of error might be the neglect of van der Waals interactions among highly polarizable Te 
atoms, absent in LDA and GGA functionals, whose inclusion might bring theoretical and experimental lattice 
parameters to a closer agreement.
The bulk modulus and its derivative  with respect to pressure obtained from the fitted Murnaghan
equation of state (PBE functional) are $B$=18.4 GPa (B=33 GPa for $\alpha$-TeO$_2$) and $B'$=18.1. 
The larger compressibility of $\beta$-TeO$_2$ with respect to $\alpha$-TeO$_2$ is due to the weak 
interlayer interaction, as demonstrated by the evolution of the lattice parameters with volume 
reported in Fig. \ref{varvol_beta}a. The $a$ axes, perpendicular to the layers, undergoes larger changes with
pressure than the other two axes do. 
However the overestimation of the lattice parameters by the PBE functional is almost the same for all the
three axes which implies that the interlayer interaction should have a strong electrostatic contribution.
Would the interlayer cohesion be simply given by van der Waals interactions, absent in GGA functional, the overestimation of
the $a$ axis by the PBE functional should have been much larger.
The change of  bond lengths and bond angles with volume is reported in Fig. \ref{varvol_beta}b-c.
By increasing the volume, the length of the shorter equatorial bonds decreases while the longer axial bonds increase in length.
This behavior would support the picture of $\beta$-TeO$_2$ as made of TeO$_2$ molecules, analogously to $\alpha$-TeO$_2$.

\begin{table}
\caption{Bond lengths and angles for $\beta\mathrm{-TeO_2}$, 
using PBE functional at the theoretical (PBE) and experimental
(PBEexp) lattice parameters, and  with BLYP functional at the experimental
 lattice parameters (BLYPexp). Experimental X-ray data (Exp) are from Ref.
 \protect\cite{beyer}. Atoms are labelled according to Fig. \protect\ref{strut_beta}.}
\label{bond_beta}
\begin{ruledtabular}
\begin{tabular}[c]{l c c c c}
                     & PBE     & PBEexp & BLYPexp  &  Exp   \\\hline
\multicolumn{5}{l}{Bond lengths (\AA)}\\\hline
Te-O1                & 1.953   & 1.960   & 1.941     & 1.876    \\
Te-O2                & 1.968   & 1.975   & 1.967     & 1.893    \\
Te-O1'               & 2.192   & 2.186   & 2.181     & 2.153    \\
Te-O2'               & 2.110   & 2.083   & 2.067     & 2.068    \\
Te-Te'\hspace{1cm}   & 3.232   & 3.240   & 3.230     & 3.168    \\\hline
\multicolumn{5}{l}{Angles (degrees)}\\\hline
O1-Te-O2             &   97.4  &  96.8   &  96.9     &   98.8   \\
O1'-Te-O2'           &  169.5  & 168.9   & 168.6     &  165.9   \\
Te-O1-Te'            &  102.3  & 102.6   & 103.0     &  103.5   \\
\end{tabular}
\end{ruledtabular}
\end{table}

\subsection{$\gamma$-TeO$_2$}

This phase has been recently identified by recrystallizing amorphous TeO$_2$ doped with oxide (Nb, W oxides, 
10 $\%$ molar). \cite{gamma1,gamma2}
The structure is orthorhombic (space group $P2_12_12_1$, $D_2^4$, n. 19) with four formula units 
and three independent atoms per cell. The theoretical (PBE) and experimental lattice parameters and the positions
of the independent atoms are reported in table  \ref{pos_gamma}.
The error in the theoretical equilibrium volume (11 $\%$) and lattice parameters (3 $\%$) is slightly larger for 
$\gamma$-TeO$_2$ than for the other phases.
The structure of $\gamma$-TeO$_2$ can still be seen as formed by corner sharing TeO$_4$ units with two-fold 
coordinated oxygen atoms and four-fold coordinated Te atoms. However, the difference in length of the four 
$\mathrm{Te-O}$ bonds is 0.34 \AA, a value sizably larger than in the other phases (0.27 \AA~ for $\alpha$-TeO$_2$ 
and 0.24 \AA~ for $\beta$-TeO$_2$).
By considering the longest $\mathrm{Te-O}$ bond (2.198 \AA) as a non-bonding interaction, $\gamma$-TeO$_2$ would 
appear as a network of TeO$_3$ units with Te atoms three-fold coordinated, two oxygen atoms two-fold coordinated and 
one non-bridging one-fold coordinated oxygen atom.

\begin{table}
\caption{Structural parameters of $\gamma\mathrm{-TeO_2}$, 
using PBE functional at the theoretical (PBE) and experimental
(PBEexp) lattice parameters, and  with BLYP functional at the experimental
 lattice parameters (BLYPexp). Experimental X-ray data (Exp) are from Ref.  
 \protect\cite{gamma2}.}
\begin{ruledtabular}
\begin{tabular}[c]{l c c c c}
                      & PBE     & PBEexp& BLYPexp &  Exp   \\\hline
\multicolumn{5}{l}{Cell parameters (\AA)}\\\hline
a                     & 5.176   &    -    &     -     & 4.898    \\
b                     & 8.797   &    -    &     -     & 8.576    \\
c                     & 4.467   &    -    &     -     & 4.351    \\\hline
\multicolumn{5}{l}{Atomic positions}\\\hline
Te $x$                & 0.9581  & 0.9797  & 0.9771    & 0.9696   \\
Te $y$                & 0.1032  & 0.1003  & 0.1007    & 0.1016   \\
Te $z$                & 0.1184  & 0.1407  & 0.1380    & 0.1358   \\
O(1) $x$              & 0.7641  & 0.7827  & 0.7801    & 0.759    \\
O(1) $y$              & 0.2851  & 0.2905  & 0.2888    & 0.281    \\
O(1) $z$              & 0.1645  & 0.1853  & 0.1805    & 0.173    \\
O(2) $x$              & 0.8599  & 0.8656  & 0.8646    & 0.855    \\
O(2) $y$              & 0.0406  & 0.0371  & 0.0361    & 0.036    \\
O(2) $z$ \hspace{1cm} & 0.7131  & 0.7237  & 0.7278    & 0.727    \\
\end{tabular}
\end{ruledtabular}
\label{pos_gamma}
\end{table}

\begin{figure}
\caption{Structure of $\gamma$-$\mathrm{TeO_2}$. Two unit cells are drawn.
(a) Three-dimensional structure. (b) Chain-like structure obtained by breaking the longest $\mathrm{Te-O}$ bond of 
the TeO$_4$ units of panel (a).  O1 and O2 ($\mathrm{O1'}$ and $\mathrm{O2'}$) label equatorial (axial) atoms.}
\includegraphics[width=1.0\columnwidth]{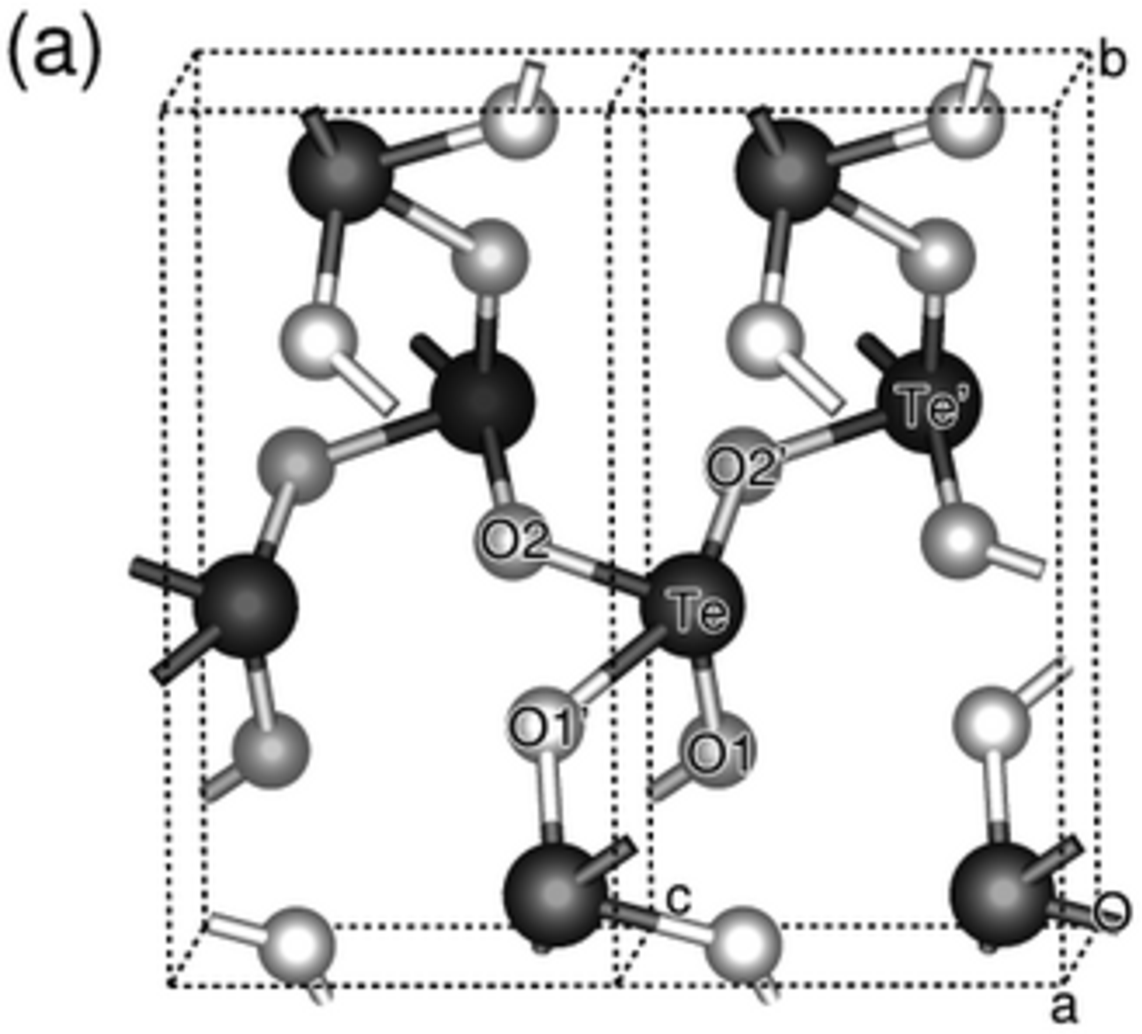}
\includegraphics[width=1.0\columnwidth]{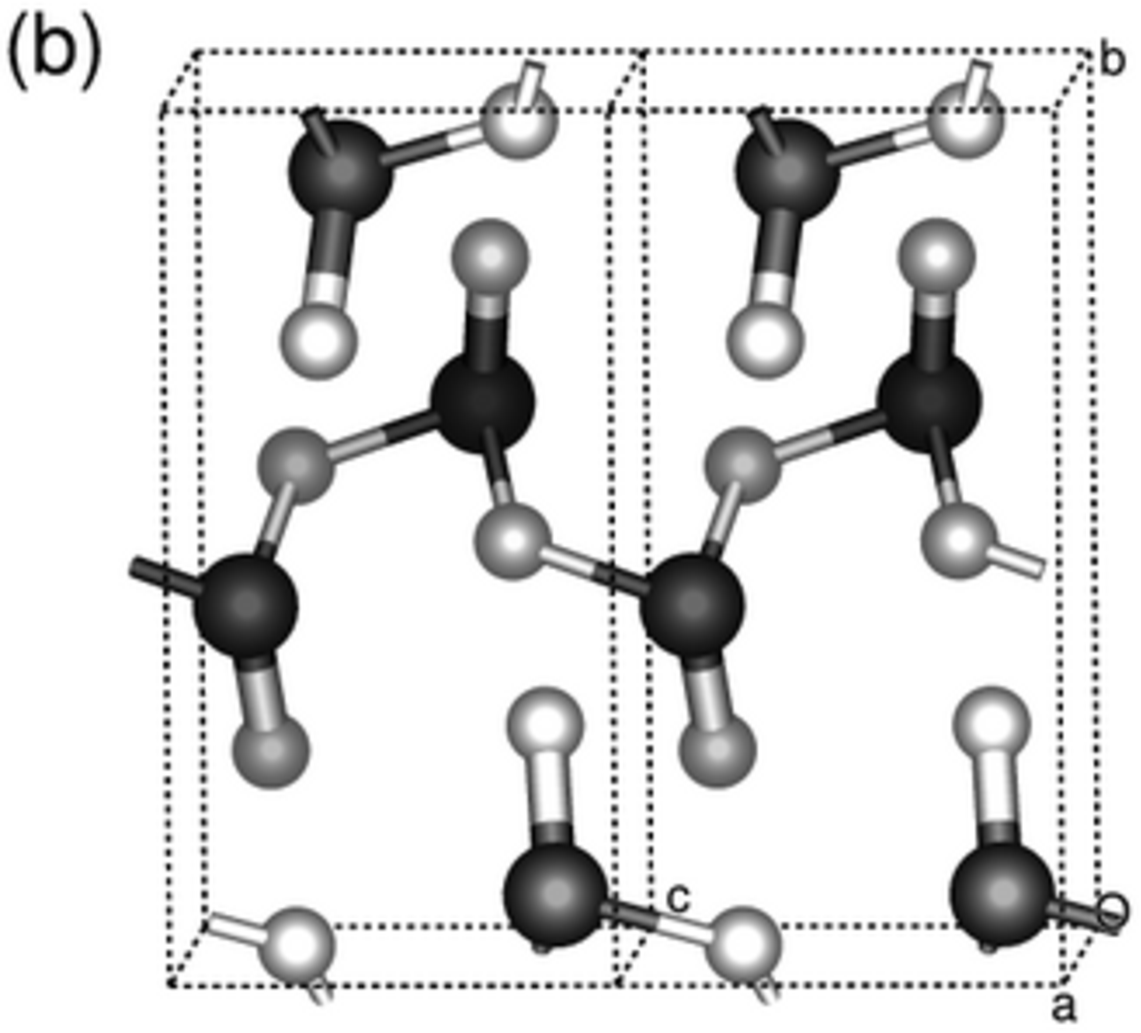}
\label{strut_gamma}
\end{figure}

\begin{figure}
\caption{(a) Lattice parameters ($a$, $b$, $c$), 
(b) bond lengths and (c) bond angles of $\gamma$-$\mathrm{TeO_2}$ as a function of unit cell
volume. The arrow indicates the theoretical equilibrium value.
 The three series in panel (a) correspond (from bottom to top) to $c$, $a$ and $b$ axes.
 The atom labels refer to Fig. \protect\ref{strut_gamma}.}
\includegraphics[width=1.0\columnwidth]{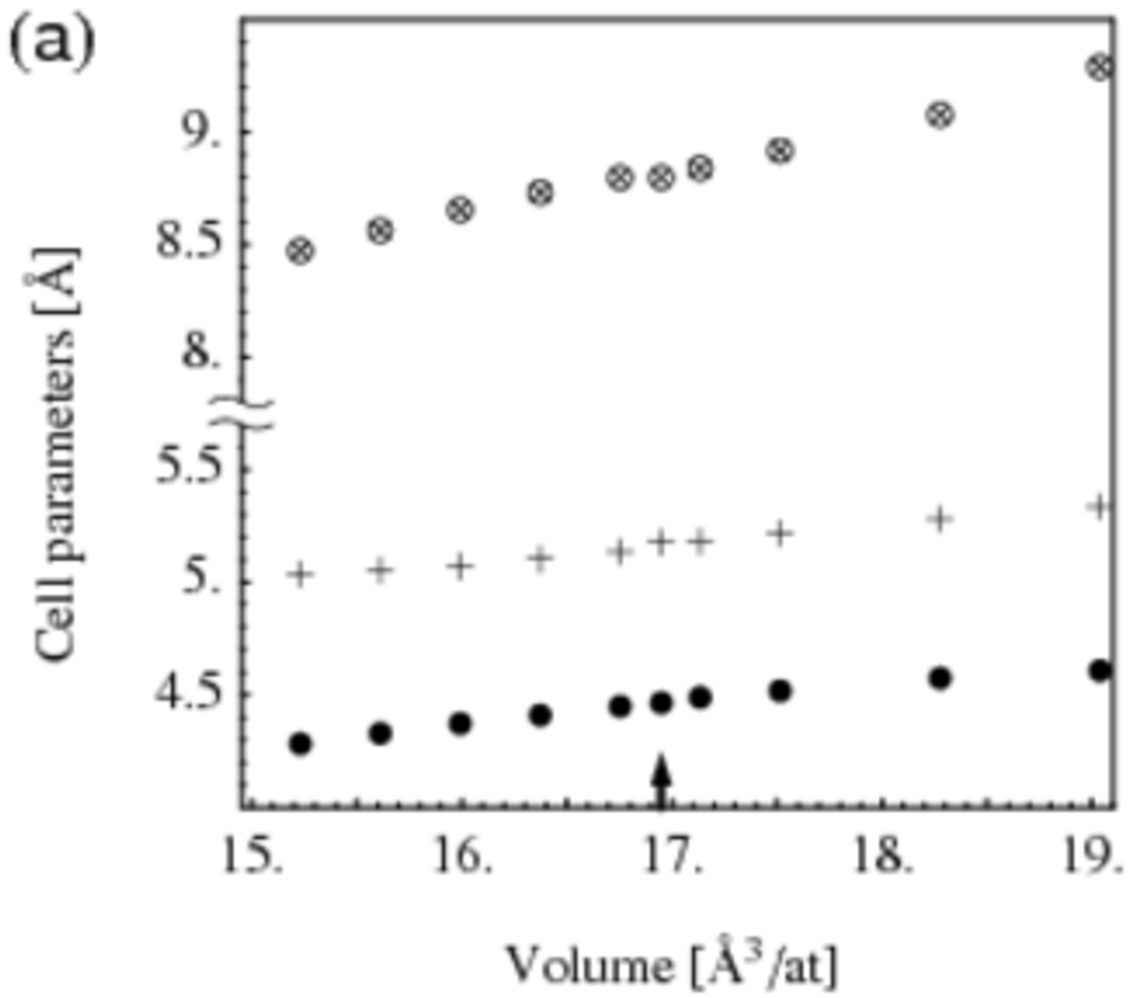}
\includegraphics[width=1.0\columnwidth]{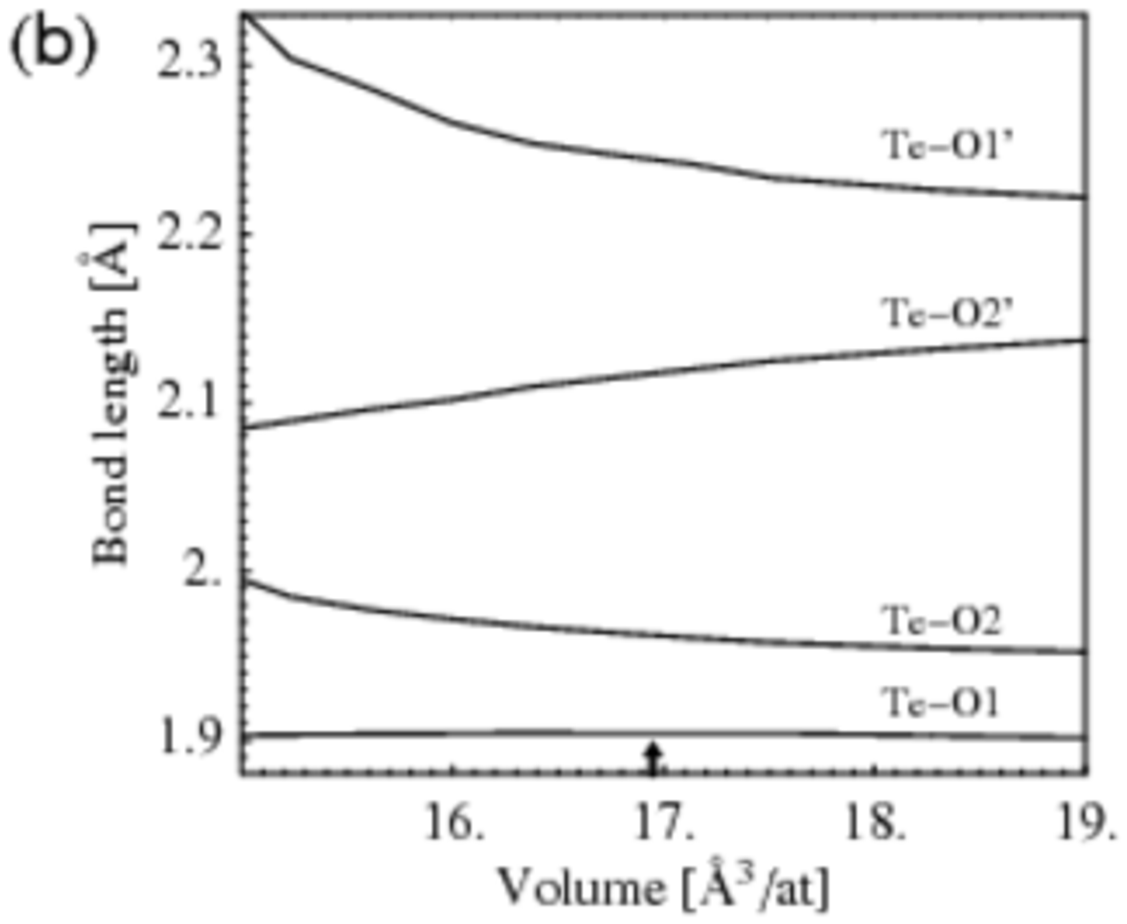}
\includegraphics[width=1.0\columnwidth]{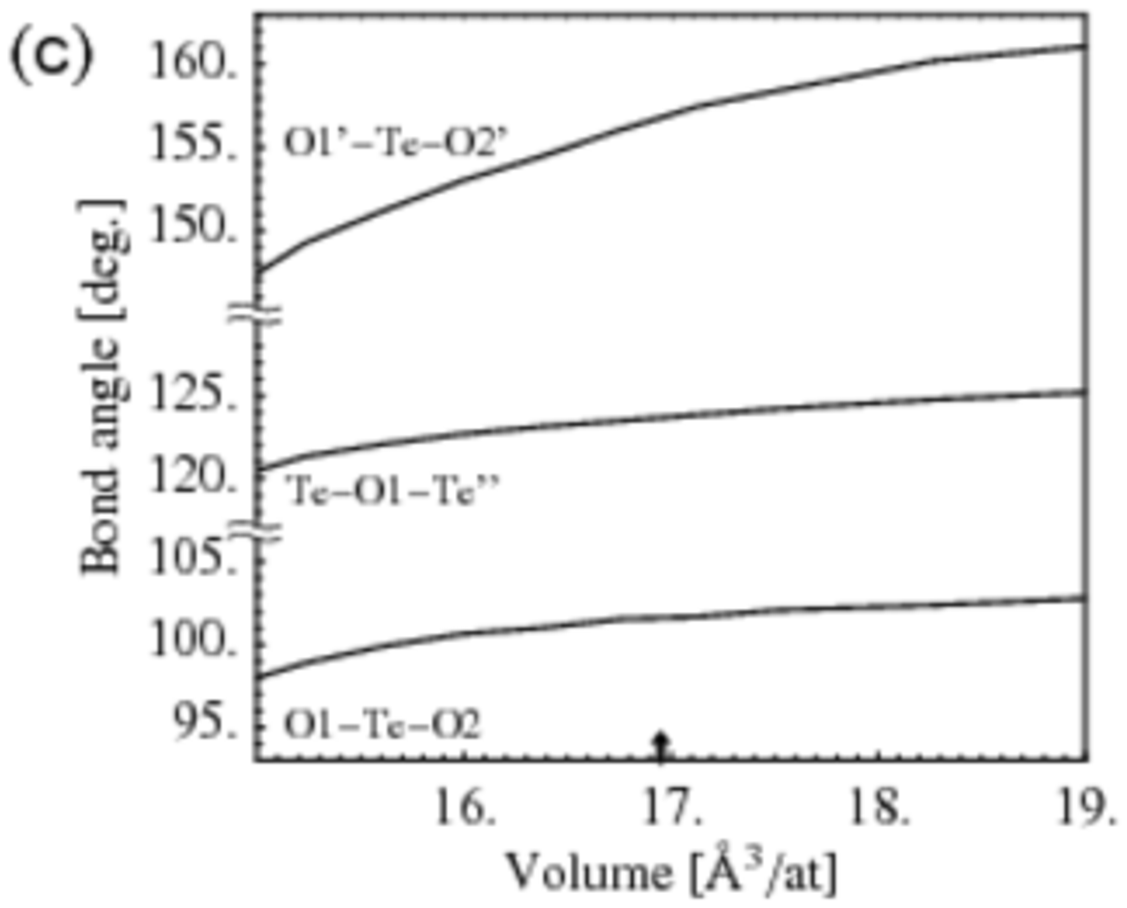}
\label{varvol_gamma}
\end{figure}

The structure of $\gamma$-TeO$_2$ depicted as a fully connected 3D network or as a one-dimensional polymeric 
structure of TeO$_3$ units is shown in Fig. \ref{strut_gamma}a and Fig. \ref{strut_gamma}b, respectively.
The calculated independent bond lengths and angles are compared with experimental data in Table \ref{bond_gamma}
for  PBE functional at the theoretical equilibrium lattice parameters and for PBE and BLYP
functionals at the experimental lattice parameters.
Still, the spread in the bond lengths is reduced in the theoretical geometry  at the theoretical lattice 
parameters (PBE) with respect to experiments. 
The larger error in the theoretical lattice parameters shows up in larger errors in the $\mathrm{Te-O}$ bond lengths in 
$\gamma$-TeO$_2$ with respect to the other phases (cfr. tables  \ref{bond_alfa} and  \ref{bond_beta}).
The misfit is largely reduced by optimizing the geometry at the experimental lattice parameters, with a 
marginal improvement for the BLYP  over PBE functional.
In particular, the optimized structure at the experimental lattice parameters reproduces the spread 
of 0.34 \AA~  in the $\mathrm{Te-O}$ bond lengths. 
The change in the lattice parameters, bond lengths and bond angles with volume is reported in Fig. \ref{varvol_gamma}.
The shorter $\mathrm{Te-O1}$ bond does not vary with pressure, while both the equatorial $\mathrm{Te-O2}$ bond and the longer $\mathrm{Te-O1'}$ axial
bond decrease in length by increasing the volume.
Regarding the picture of $\gamma$-TeO$_2$ as a polymeric phase,
 we note that the longest $\mathrm{Te-O}$ bond  of the TeO$_4$ unit
(2.198 \AA) is only 5 $\%$ longer than the other axial bond. 
On the other hand, one notes that in $\gamma$-TeO$_2$
the difference in length between the longer equatorial bond and the short axial bond is the smallest among the
three crystalline phases. The polymeric structure  would then be seen as generated from the formation of 
strong intermolecular bonds between the TeO$_2$ structural units of  tellurite and paratellurite rather than from the
breaking of one bond of the TeO$_4$ units.  \cite{gamma1,gamma2,fononi_gamma}
The one-dimensional character of $\gamma$-TeO$_2$ has to be supported by 
the observation of strong anisotropies in the
physical properties of this phase.
To address this issue, we have calculated the elastic constants of $\gamma$-TeO$_2$  from the difference in
the total energy of strained crystals. A maximum strain of 3 $\%$ has been considered.
 For stretching along the principal axes we
obtain $C_{11}$=38 GPa, $C_{22}$=41 GPa, and $C_{33}$=43 GPa, where the Voight notation has been used.
The bulk modulus and its derivative  with respect to pressure obtained from the fitted Murnaghan
equation of state (PBE functional) are $B$=16 GPa ($B$=33 GPa for $\alpha$-TeO$_2$) and $B'$=6.2.
The elastic properties are thus nearly isotropic and do not support the picture of  $\gamma$-TeO$_2$ as 
made of  one-dimensional polymeric structure.
The  elastic constants reported above differ from the results  of  calculations 
with empirical interatomic potentials ($C_{11}$=30 GPa, $C_{22}$=40 GPa, and $C_{33}$=67 GPa,
Ref. \cite{gamma2}), mainly in the $C_{33}$ component which shows an appreciable lower compressibility
along the chain ($z$) axis, not confirmed by our ab-initio  results. 

\begin{table}
\caption{Bond lengths and angles for $\gamma\mathrm{-TeO_2}$, 
using PBE functional at the theoretical (PBE) and experimental
(PBEexp) lattice parameters, and  with BLYP functional at the experimental
 lattice parameters (BLYPexp). Experimental X-ray data (Exp) are from Ref.  
 \protect\cite{gamma2}.  Atoms are labelled according to Fig. \protect\ref{strut_gamma}.}
\label{bond_gamma}
\begin{ruledtabular}
\begin{tabular}[c]{l c c c c}
                     & PBE     & PBEexp& BLYPexp &  Exp   \\\hline
\multicolumn{5}{l}{Bond lengths (\AA)}\\\hline
Te-O1                & 1.900   & 1.905   & 1.889     & 1.859    \\
Te-O2                & 1.960   & 1.974   & 1.949     & 1.949    \\
Te-O1'               & 2.252   & 2.257   & 2.240     & 2.198    \\
Te-O2'               & 2.119   & 2.092   & 2.081     & 2.019    \\
Te-Te'\hspace{1cm}   & 3.596   & 3.571   & 3.559     & 3.521    \\\hline
\multicolumn{5}{l}{Angles (degrees)}\\\hline
O1-Te-O2             &   99.1  & 100.7   & 100.9     &  101.5   \\
O1'-Te-O2'           &  156.8  & 150.9   & 152.5     &  153.6   \\
Te-O1-Te'            &  123.6  & 122.8   & 124.0     &  125.0   \\
\end{tabular}
\end{ruledtabular}
\end{table}

\begin{figure}
\caption{Electronic band structure of (a) $\beta$-$\mathrm{TeO_2}$ and (b) $\gamma$-$\mathrm{TeO_2}$. 
The orthorhombic irreducible 
 Brillouin zone is reported in the inset, equal in shape (but not in size) for both crystals.}
\includegraphics[width=1.0\columnwidth]{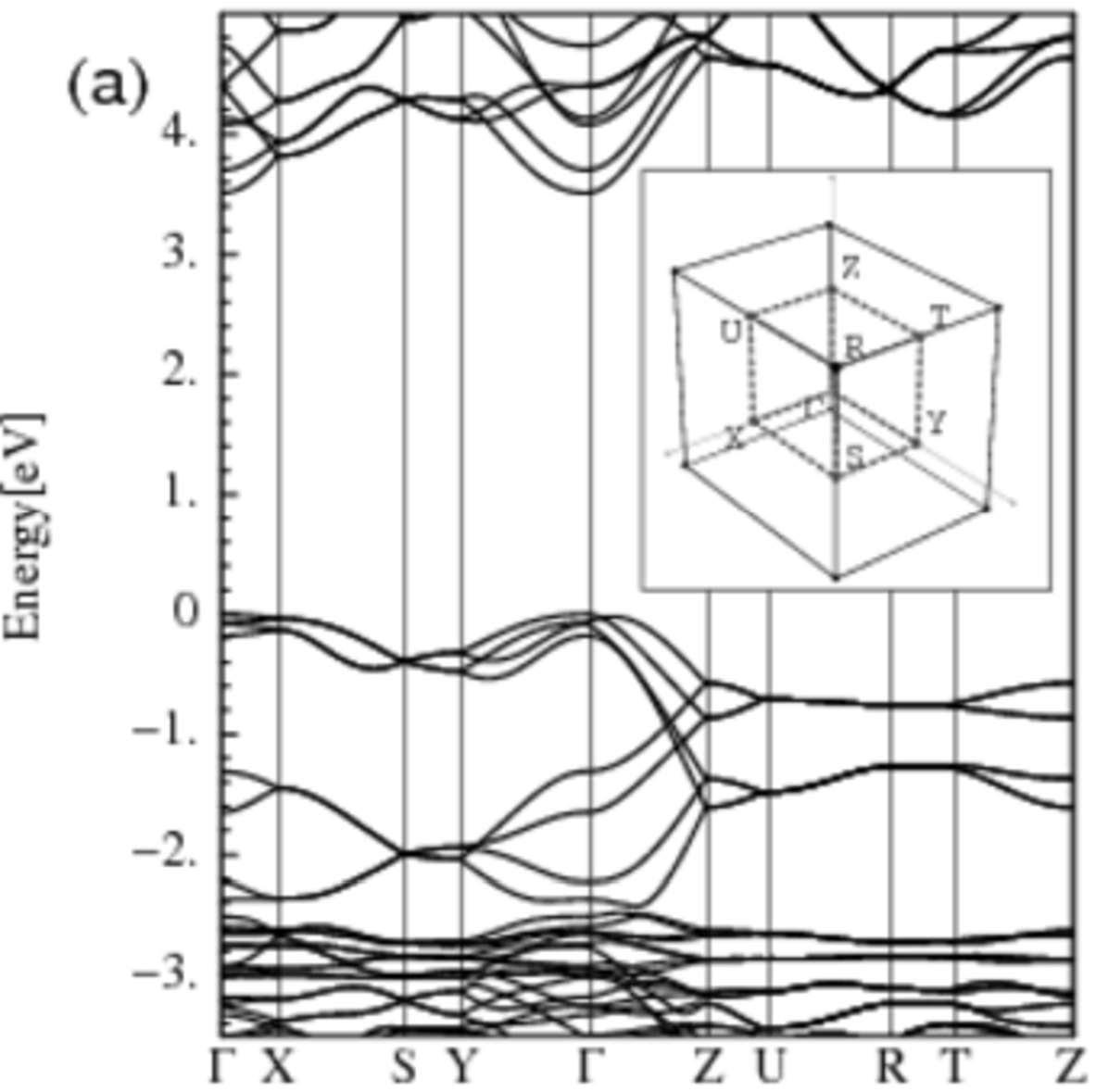}
\includegraphics[width=1.0\columnwidth]{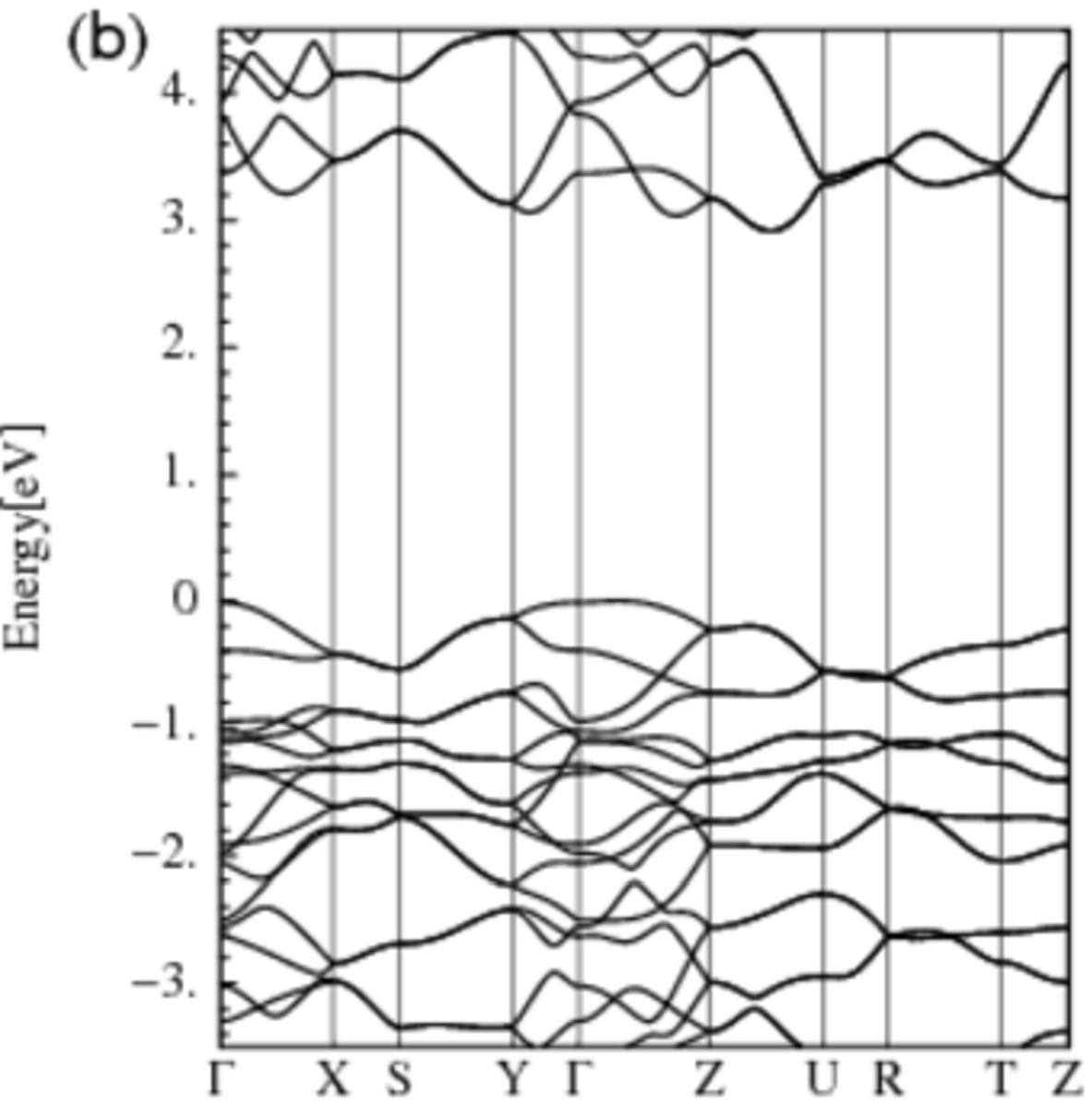}
\label{bande}
\end{figure}

We have also contrasted
the electronic band structure of $\gamma$-TeO$_2$ and
$\beta$-TeO$_2$ phases in Fig. \ref{bande}.
The electronic band structure of $\alpha$-TeO$_2$ has been computed within DFT in  a previous work.
\cite{kleinman}
In $\beta$-TeO$_2$, the band dispersion along the direction  $\Gamma$-$X$, perpendicular to the 2D layers,
is sizably smaller than along the $\Gamma$-$Y$ and $\Gamma$-$Z$ directions within  the layer plane.
Conversely, in $\gamma$-TeO$_2$, the dispersion is similar along the three orthogonal directions $\Gamma$-$X$,
$\Gamma$-$Y$, $\Gamma$-$Z$. Nor the dispersion along $\Gamma$-$Z$ is much larger in $\gamma$-TeO$_2$ than
in $\beta$-TeO$_2$, as one would expect by comparing a polymeric phase with chains along $z$ ($\gamma$ phase)
with a mainly molecular crystal made of TeO$_2$ molecules ($\beta$ phase).
Therefore, the electronic band structure does not provide  evidence of a  polymeric  nature of
$\gamma$-TeO$_2$.

\begin{figure}
\caption{Angular dispersion of  E and A$_2$
IR-active modes in $\alpha$-$\mathrm{TeO_2}$. Dashed lines represent unshifted (i.e. TO) $E$ modes.}
\includegraphics[width=1.0\columnwidth]{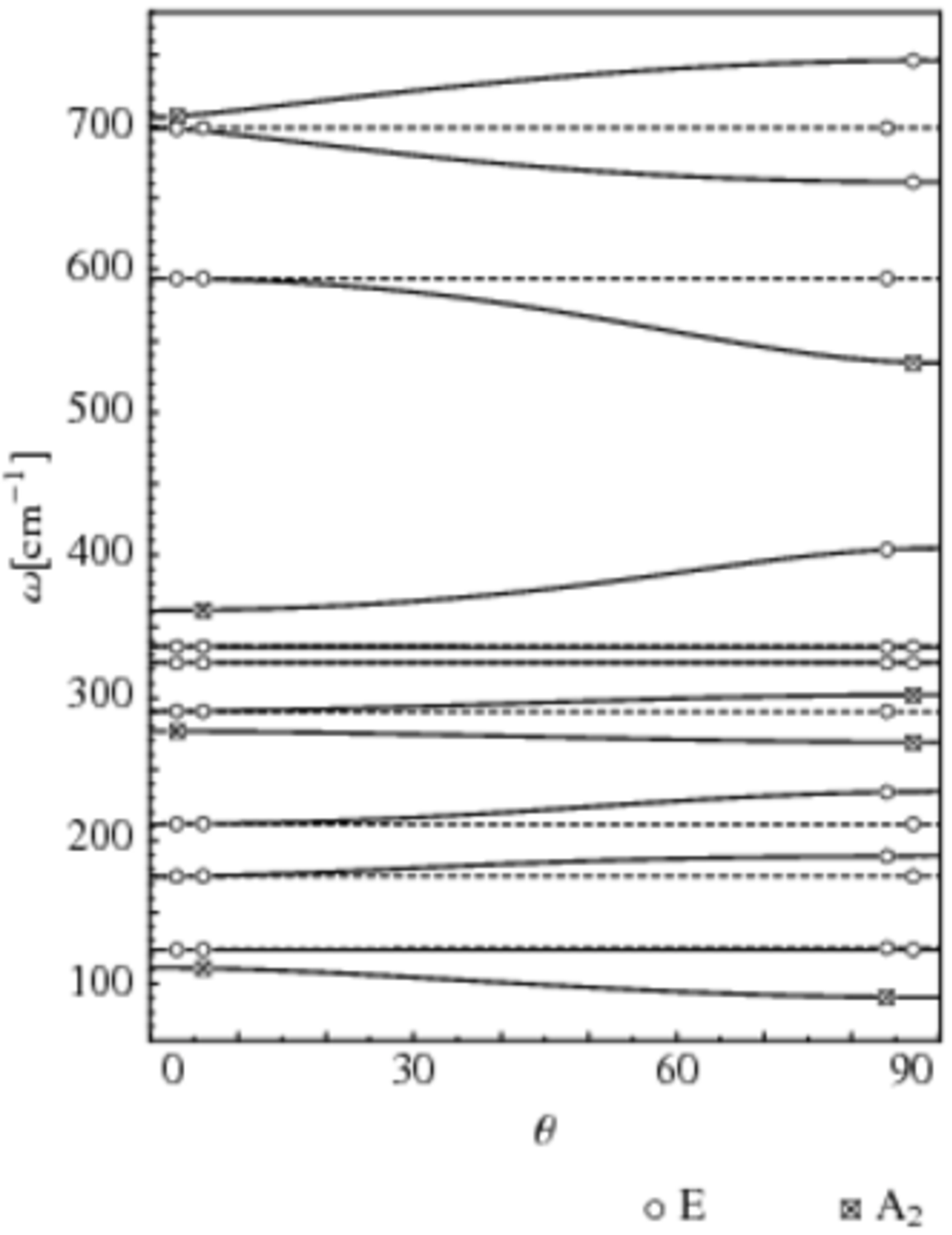}
\label{disp_alfa}
\end{figure}

We have also computed formation enthalpy and formation free energy at normal conditions
(298.15 K, p=1 bar) for the three cristalline  phases, by including the phononic 
contribution to free energy and enthalpy.
Only $\Gamma$-point phonons have been included in the calculation 
(see section \ref{vibra}). For $\alpha$-$\mathrm{TeO_2}$ we have verified
 that the phononic contribution to the free energy of -7.8 kJ/mol, 
which includes only $\Gamma$-point  phonons, changes to -6.7 kJ/mol by including
in the calculation the full phonon density of states.
Formation free energies and enthalpies have been computed with PBE functionals 
at the theoretical equilibrium structure for all the three phases.

The calculated formation enthalpy ($\Delta H_f$) and free energy ($\Delta G_f$) of 
$\alpha$-$\mathrm{TeO_2}$, with respect to 
crystalline Te and gaseous O$_2$  (the standard states), are 
$\Delta H_f=$-287.5 kJ/mol (exp. -322.6 kJ/mol)\cite{handbook}, 
and $\Delta G_f=$-232.8 kJ/mol (exp. -270.3 kJ/mol)\cite{handbook}. The 
error is within the usual accuracy of DFT methods.
The free energies of gaseous Te$_2$ and O$_2$ have been calculated from
the molecule optimized in the triplet state,\cite{o2te2} by including translational, 
vibrational and rotational contributions. The free energy of crystalline
Te has been computed from theoretical and experimental\cite{handbook} free
energies of the gaseous Te$_2$ molecule.

For $\beta$-$\mathrm{TeO_2}$ the calculated formation enthalpy and free energy are
$\Delta H_f=$-289.3 kJ/mol and  $\Delta G_f=$-236.3 kJ/mol, while for $\gamma$-$\mathrm{TeO_2}$
$\Delta H_f=$-284.6 kJ/mol and  $\Delta G_f=$-229.8 kJ/mol. To our knowledge,
no experimental data on formation energies are available for $\beta$-$\mathrm{TeO_2}$
and $\gamma$-$\mathrm{TeO_2}$.

\section{\label{vibra}Vibrational properties}

Phonons  at the $\Gamma$-point have been calculated within density functional
perturbation theory.\cite{dfpt} 
The  modes which display a dipole moment  couple to the inner macroscopic longitudinal electric field which 
shifts the LO phonon frequencies via the non-analytic contribution to the dynamical matrix \cite{dfpt}

\begin{equation}  
D^{NA}_{\alpha\beta}(\kappa,\kappa')=\frac{4 \pi}{V_o} \frac{Z_{\alpha\alpha'}(\kappa)q_{\alpha'}
Z_{\beta\beta'}(\kappa')q_{\beta'} }
{{\bf q} \cdot \tens{\epsilon}^{\infty}  \cdot {\bf q}},
\label{macro}
\end{equation}

where $\tens{Z}$ and $\tens{\epsilon}^{\infty}$ are the effective charges and electronic dielectric tensors, 
$V_o$ is the unit cell volume and {\bf q} is the phononic wavevector.
The macroscopic field contribution to the dynamical matrix (Eq. \ref{macro}) introduces an angular dispersion 
of the phonons at the $\Gamma$ point, i.e. the limit of the phononic bands $\omega({\bf q})$ for ${\bf q} 
\rightarrow 0$ depends on the angles formed by {\bf q} with the principal axis.
For a uniaxial crystal as $\alpha$-TeO$_2$, the phonon frequencies depend on the angle $\theta$ formed by {\bf q} 
with the optical axis. For the orthorhombic crystals $\beta$-TeO$_2$ and $\gamma$-TeO$_2$, the phonon frequencies
depend on the polar angles $\theta$ and $\phi$ which define the orientation of {\bf q} with respect to the
crystallographic axes,  coinciding with the optical principal axes.

The dielectric tensor is given in terms of phonons and effective charges by

\begin{equation}
\epsilon_{\alpha \beta}(\omega)  = \epsilon^{\infty}_{\alpha \beta} + \frac{4 \pi}{V_o}\sum_{j=1}^{3N} 
p_{\alpha}(j)p_{\beta}(j)\frac{1}{\omega_j^2-\omega^2} \\
\label{epsi}
\end{equation}

where $\mathbf{p}(j)$ is given by

\begin{equation}
p_{\alpha}(j)=\sum_{\kappa=1}^N {Z}_{\alpha\beta}(\kappa) \frac{e_{\beta}(j,\kappa)}{\sqrt{M_{\kappa}}}.
\\
\label{dipole}
\end{equation}

The absorption coefficient $\alpha_{\gamma}$ for light polarized along the crystallographic axis $\gamma$ is

\begin{eqnarray}
\alpha_{\gamma}(\omega) &  = & \frac{\omega}{nc} {\rm Im}~\epsilon_{\gamma \gamma}
(\omega+i\eta, \eta \rightarrow 0) \nonumber \\
 &  =  &  \frac{2 \pi^2}{V_onc}\sum_{j} \left| p_{\gamma}(j)\right|^2 \delta(\omega-\omega_j) \nonumber\\
 &  =  &  \frac{ \pi}{2 nc}\sum_{j} f_j \omega^2_j \delta(\omega-\omega_j), \label{absorb}
\end{eqnarray}

where $c$ is the velocity of light in vacuum and $n(\omega)$ is the (frequency dependent) real part of the 
refractive index.

In order to compare the theoretical results with available experimental data we have also computed the
IR absorption for a polycrystalline sample. This amounts to average the absorption coefficient for the
two  polarizations of the electromagnetic wave (ordinary and extraordinary waves for uniaxial crystals, for instance)
over  the solid angle of the possible wave vectors {\bf q} of the transmitted wave.
The absorption coefficient for a generic {\bf q} is obtained by solving the Fresnel's equations for the
dielectric tensor defined by Eq. \ref{epsi}. \cite{claus}

The differential cross section for Raman scattering (Stokes) in non-resonant conditions
is given by the following expression (for a unit volume of scattering sample)
\begin{equation}
\frac{d^2 \sigma}{d \Omega d \omega} = \sum_{j} \frac{\omega_S^4 }{c^4} \left| {\bf e}_S 
\cdot \tens{R}^j \cdot  {\bf e}_L\right|^2 (n_B(\omega)+1)\delta(\omega-\omega_j),
\label{raman}
\end{equation}
where  $n_B(\omega)$ is the Bose factor, $\omega_S$ is the frequency of the scattered light, ${\bf e}_S$ 
and ${\bf e}_L$ are the polarization vectors of the scattered and incident light, respectively.  \cite{cardona,bruesh}
The Raman tensor $\tens{R}^j$ associated with the $j$-th phonon is given by
\begin{equation}
R_{\alpha \beta}^j =  \sqrt{\frac{V_o \hbar}{2 \omega_j}}
\sum_{\kappa=1 }^N \frac{\partial \chi_{\alpha \beta}^{\infty}}{\partial {\bf r}(\kappa)}
\cdot \frac{{\bf e}(j,\kappa)}{\sqrt{M_{\kappa}}},
\label{ramanT}
\end{equation}
where $V_o$ is the unit cell volume, ${\bf r}(\kappa)$ is the position of the $\kappa$-th atom and 
$\tens{\chi}^{\infty}=(\tens{\epsilon}^{\infty}-{\mathbbm{1}})/4\pi$ is the electronic susceptibility.
The tensor $\tens{R}^j$ is computed from $\tens{\chi}^{\infty}$ by finite differences, by moving the atoms 
independent by symmetry  with maximum displacement of 0.01 \AA.

Whenever the experimental Raman spectrum is available only for  a polycrystalline sample, Eq. \ref{raman} 
must be integrated over the solid angle in order to compare the theoretical spectra with experiments.
In particular, the Raman spectrum for  non-polarized light  is obtained by summing over all possible 
polarization vectors  ${\bf e}_S$ and ${\bf e}_L$ consistent with the scattering geometry.
In the case of $\beta$-TeO$_2$, the inner longitudinal macroscopic field has no effect on the Raman active 
modes due to the presence of an inversion symmetry. In fact, the Raman active $g$-modes have no dipole moment 
to couple with the macroscopic longitudinal field.
As a consequence for $\beta$-TeO$_2$ there is no angular dispersion for the Raman-active modes and 
the  total cross section for unpolarized light in backscattering geometry is obtained from Eq. \ref{raman}
with the substitution

\begin{eqnarray}
4(R_{xx}^2+R_{yy}^2+R_{zz}^2) +7(R_{xy}^2+R_{xz}^2+R_{yz}^2)+ \nonumber \\
(R_{xx}R_{yy}+R_{xx}R_{zz}+R_{zz}R_{yy})
\rightarrow  30 | {\bf e}_S \cdot \tens{R}^j \cdot  {\bf e}_L|^2 \label{solidangle}
\end{eqnarray}

Conversely,  for $\alpha$-TeO$_2$ and  $\gamma$-TeO$_2$ the presence of  angular dispersion requires 
the integral over the solid angle to be performed by summing over discrete points in the polar angles. 
Phonon frequencies and eigenvectors entering in Eq. \ref{raman} and  \ref{ramanT} are modified by the 
macroscopic longitudinal field  along the direction assigned by the crystal momentum transferred in the 
scattering process.

The $\delta$-functions in Eq. \ref{absorb} and \ref{raman} are approximated by Lorentzian functions as

\begin{equation}    
\delta(\omega-\omega_j)=\frac{4}{\pi}\frac{\omega^2 \eta_j}{(\omega^2-\omega_j^2)^2 + 4 \eta_j^2\omega^2}.
\label{lorenz}
\end{equation}

with $\eta_j$ fitted on the experimental Raman peaks when available and otherwise assigned to a constant value
of 4 cm$^{-1}$ (cfr. Table \ref{phon_alfa}). 

\subsection{$\alpha$-TeO$_2$}

Phonons at the $\Gamma$-point can be classified according to the irreducible representations of the 
$D_4$ point group of $\alpha$-TeO$_2$ as $\Gamma=4A_1+4A_2+5B_1+4B_2+8E$, where 
the acoustic modes ($A_2$ and $E$) have been omitted. 
The $A_1$, $B_1$, $B_2$ and $E$ modes and Raman active, while the $A_2$ and $E$ modes are IR active.
The calculated phonon frequencies at the $\Gamma$ point neglecting the contribution of the macroscopic 
longitudinal field are given in Table \ref{phon_alfa} for the BLYP calculation at the experimental lattice 
parameters. 
The angular dispersion of $A_2$ and $E$ modes due to the macroscopic electric field is 
reported in Fig. \ref{disp_alfa}; it is in good agreement with the angular dispersion 
obtained experimentally from the IR absorption peaks of the extraordinary waves 
(cfr. Fig. 3 of Ref. \cite{expIRalfa}).
Experimental frequencies and IR activities  of IR-active modes 
($A_2$ and $E$(TO) modes) are also reported in Table \ref{phon_alfa}.
Experimental frequencies from Raman spectra are also reported in Table \ref{phon_alfa}
for $A_1$, $B_1$ and $B_2$ modes.
For Raman active $E$(LO) modes the comparison is made directly 
between the theoretical and experimental Raman spectra, since
their frequencies  depend on the scattering geometries. 
Phonon frequencies calculated with PBE or LDA functional at the theoretical lattice parameters and with PBE and BLYP 
functionals at the  experimental lattice parameters are compared in Table 
 \ref{phon_alfa}.
The best agreement with experiments is obtained for BLYP calculations at the experimental lattice
parameters  which give the equilibrium internal geometry closest to the experimental structure
(cfr. section IIIA).
The IR oscillator strengths are in good agreement with experiments, apart from the overestimation of the
intensity of mode $A_2$(2). 
Experimental data from a more recent work \cite{twoIR} are very similar (within 5 cm$^{-1}$) to
 the older experimental data reported
in Table \ref{phon_alfa}, but for the oscillator strength ($f$) of mode $A_2$(2) which is 3.33 in Ref. \cite{twoIR},
in closer agreement with our theoretical result.
The IR absorption spectrum for a polycrystalline sample, computed as described in the previous section 
is reported in Fig. \ref{poly_ir}; it has to be compared with the experimental spectrum reported in Fig. 2 of Ref. \cite{fononi_beta}.
The linewidth of the phononic modes is chosen according to the IR experimental data on single crystal. \cite{expIRalfa}
The two broad bands and the shoulder at lower  frequency in the experimental spectrum\cite{fononi_beta} are well reproduced, but
for the already mentioned redshift in the theoretical frequencies.

\begin{table}
\caption{
 Theoretical phonon frequencies of $\alpha$-TeO$_2$ at the $\Gamma$ point, oscillator strengths
 ($f_j$ in Eq. \protect\ref{absorb}) of IR active modes and squared coefficients of the Raman
 tensor of the Raman active modes, $a^2$ and $b^2$ for $A_1$, $c^2$ for $B_1$, $d^2$
 for $B_2$, and $e^2$ for $E_x$, $E_y$ modes (in units of $10^{-3}$\AA$^3$, see section IV).
 The contribution of the inner longitudinal macroscopic field is not included (LO-TO splitting).
 All the theoretical values correspond to calculations with the BLYP functional at the
 experimental lattice parameters. 
 Experimental  data  on phonon frequencies and oscillator strengths from IR absorption spectra
at 85 K \protect\cite{expIRalfa} are given in parenthesis for (TO) $A_2$ and $E$  modes.
Experimental phonon frequencies of $A_1$, $B_1$ and $B_2$ modes (in parenthesis) are taken from
Raman spectra at 85 K \protect\cite{Pine}. $\Delta\omega$ indicate the phonon linewidths obtained by
fitting the main peaks of the polycrystalline Raman spectra at 295 K (cfr. Fig.  \protect\ref{raman_blyp_exp}).}
\begin{ruledtabular}
 \begin{center}
  \begin{tabular}{l c c c c c c c}
 Mode & \multicolumn{2}{c}{$\omega$ (cm$^{-1}$)} & \multicolumn{2}{c}{$f_j$} & $a^2$ ($c^2$, $d^2$, $e^2$) & $b^2$ & $\Delta\omega$\\
   \hline
 $B_1$ (1)          &    42 &  (62)   &         &         &  2.044 &       &      \\
 $A_2$ (1)          &    90 &  (76)   &  7.758  & (12.95) &        &       &      \\
 $E$   (1)          &   124 &  (124)  &  0.919  & (1.28)  &  5.158 &       &  2.9 \\
 $B_1$ (2)          &   128 &         &         &         &  0.042 &       &      \\
 $A_1$ (1)          &   143 &  (152)  &         &         &  4.939 &  7.561&  5.8 \\
 $B_2$ (1)          &   148 &  (157)  &         &         &  2.767 &       &  2.2 \\
 $B_1$ (3)          &   175 &  (179)  &         &         &  0.324 &       &      \\
 $E$   (2)          &   175 &  (177)  &  5.667  & (8.05)  &  0.980 &       &      \\
 $E$   (3)          &   212 &  (212)  &  3.127  & (2.02)  &  0.043 &       &      \\
 $A_1$ (2)          &   214 &  (218)  &         &         &  0.338 &  0.058&      \\
 $B_1$ (4)          &   229 &  (235)  &         &         &  0.514 &       &      \\
 $A_2$ (2)          &   269 &  (265)  &  3.156  & (0.77)  &        &       &      \\
 $B_2$ (2)          &   272 &  (281)  &         &         &  0.330 &       &      \\
 $E$   (4)          &   291 &  (299)  &  5.121  & (3.95)  &  0.045 &       &      \\
 $A_2$ (3)          &   302 &  (325)  &  4.096  & (4.70)  &        &       &      \\
 $E$   (5)          &   325 &  (336)  &  0.005  & (0.25)  &  0.077 &       &      \\
 $E$   (6)          &   336 &  (379)  &  0.006  & (0.01)  &  0.863 &       &  6.4 \\
 $A_1$ (3)          &   383 &  (391)  &         &         &  2.421 &  7.923& 12.1 \\
 $B_2$ (3)          &   406 &  (415)  &         &         &  0.136 &       &      \\
 $A_2$ (4)          &   535 &  (570)  &  3.802  & (3.86)  &        &       &      \\
 $B_1$ (5)          &   537 &  (589)  &         &         &  1.551 &       & 14.0 \\
 $E$   (7)          &   594 &  (644)  &  1.268  & (1.27)  &  0.200 &       &      \\
 $A_1$ (4)          &   598 &  (649)  &         &         & 10.242 & 14.815&  4.7 \\
 $E$   (8)          &   700 &  (774)  &  0.225  & (0.21)  &  0.671 &       &      \\
 $B_2$ (4)          &   726 &  (786)  &         &         &  0.574 &       &      \\
  \end{tabular}
 \end{center}
\end{ruledtabular}
\label{phon_alfa}
\end{table}

The Raman tensor (Eq. \ref{ramanT}) for the Raman-active  irreducible representations has the following form:\cite{loudon}

\begin{displaymath}
\begin{array}{rr}
A_1 \Rightarrow
\left [ 
\begin{array}{ccc}
   a  &  ~.~ &  ~.~ \\
  ~.~ &   a  &  ~.~ \\
  ~.~ &  ~.~ &   b  \\
\end{array}
\right ]
\label{a1}
&
B_1 \Rightarrow
\left [ 
\begin{array}{ccc}
   c  &  ~.~ &  ~.~ \\
  ~.~ &  -c  &  ~.~ \\
  ~.~ &  ~.~ &  ~.~ \\
\end{array}
\right ]
\label{b1}
\\ & \\
B_2 \Rightarrow
\left [ 
\begin{array}{ccc}
  ~.~ &   d  &  ~.~ \\
   d  &  ~.~ &  ~.~ \\
  ~.~ &  ~.~ &  ~.~ \\
\end{array}
\right ]
\label{b2}
&\rule{5mm}{0pt}
E_x(y) \Rightarrow
\left [ 
\begin{array}{ccc}
  ~.~ &  ~.~ &  (-e) \\
  ~.~ &  ~.~ &   e   \\
  (-e) &   e  &  ~.~ \\
\end{array}
\right ]
\label{ex}
\\\end{array}
\end{displaymath}

The coefficients $a$,$b$,$c$,$d$, and $e$ calculated from first principles as outlined above are given 
for each mode in Table \ref{phon_alfa}.
The calculated derivatives of the dielectric tensor with respect to the displacement
of the atoms independent by symmetry are given as additional materials.\cite{epaps}

\begin{figure}
\caption{Raman spectra of  $\alpha$-$\mathrm{TeO_2}$ single crystal for different scattering geometries.  Theoretical spectra (th.)
 are computed with the BLYP functional at experimental lattice parameters. 
The experimental spectra at 85 K are adapted from Ref. \protect\cite{Pine}.
The character of the modes active in each scattering geometry is also reported.
The linewidths of the theoretical  peaks are fitted on the experimental spectra whenever the correspondence is unambiguous.
The Raman intensities are reported on a logarithmic scale for sake of comparison with the experimental data 
of Ref. \protect\cite{Pine}. The experimental peak is normalized to the strongest peak of the corresponding theoretical spectrum.}
\vspace{1.5cm}
\includegraphics[width=1.0\columnwidth]{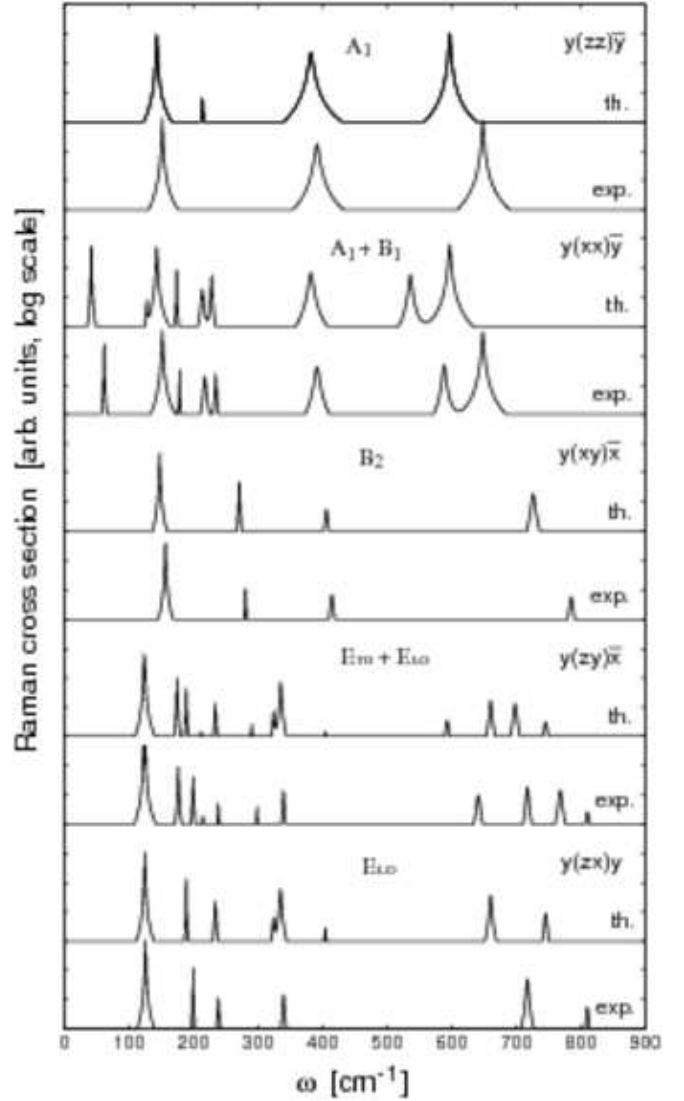}
\label{raman_alfa}
\end{figure}

\begin{figure}
\caption{Phonon dispersion relations and density of states of $\alpha$-$\mathrm{TeO_2}$,
 calculated with  BLYP functional at the experimental lattice parameters.}
\includegraphics[width=1.0\columnwidth]{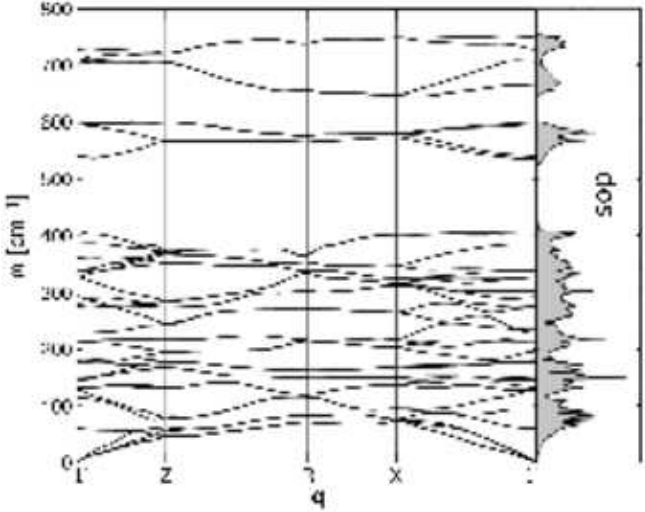}
\label{dos_alfa}
\end{figure}

\begin{figure}
\caption{IR absorption ($\alpha(\omega)$, cfr. Eq. \protect\ref{absorb}) 
spectra for polycrystalline samples of $\alpha$-, $\beta$-, and $\gamma$-$\mathrm{TeO_2}$; linewidths in $\alpha$-$\mathrm{TeO_2}$ are 
taken from experimental data on single crystals\cite{expIRalfa}. In the lack of experimental data, the linewidths   $\beta$-, and 
$\gamma$-$\mathrm{TeO_2}$ are chosen  analogously to those of  $\alpha$ phase, i.e. 5 $cm^{-1}$ 
for modes with frequency lower than 150 $cm^{-1}$, and 15 $cm^{-1}$ for higher frequency modes.}
\includegraphics[width=1.0\columnwidth]{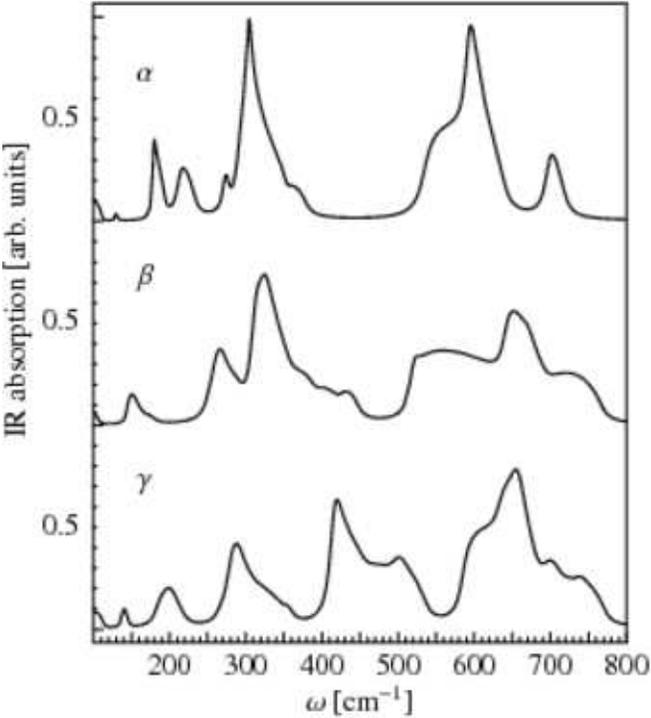}
\label{poly_ir}
\end{figure}

The theoretical Raman spectra for different scattering conditions, each selecting particular modes, are
compared with experimental data,\cite{Pine,gamma2} in  Fig. \ref{raman_alfa}.
The theoretical spectra include
the shift in phonon frequencies due to the contribution of the macroscopic longitudinal field along
the direction assigned by the crystal momentum transferred in the scattering geometry.
The Raman spectrum for a polycrystalline sample is also computed and compared with experimental data 
in Fig. \ref{raman_blyp_exp}. 
For each wavevector {\bf q} transfered in the scattering process the phonon eigenvector and frequency  
is computed by including the non-analytic part of the dynamical matrix.
The solid angle average is performed over 256 angles independent by symmetry. 
The linewidth of the most intense Raman peaks in the theoretical spectra has been obtained by fitting the experimental
spectra with a sum of Lorentzian functions, either at 85 K for the single crystal spectra (Fig. \ref{raman_alfa})
or at 295 K for the polycrystalline spectra. The high temperature linewidth are given in Table \ref{phon_alfa}.
The linewidth of the weaker or less identifiable peaks is set to 4 cm$^{-1}$.

The calculated Raman spectra are in good agreement with experiments 
on peak intensities and also on peak frequencies
in the low-frequency region, while the
high-frequency modes appear too soft by nearly 8\% (see Table \ref{phon_alfa}). The discrepancy can possibly be
ascribed to the partial homogenization of axial and equatorial bonds resulting from the use of approximate 
exchange-correlation functionals. 
This effect is smaller for BLYP  than for PBE functionals (cfr. section IIIA).
The experimental Raman peak at 218 cm$^{-1}$ assigned to a $B_1$ mode in Ref. \cite{Pine} can be better
assigned to the $A_1$(2) mode on the base of our calculations (cfr. Table \ref{phon_alfa}) which is
also not inconsistent with experimental two-phonons IR spectra of Ref. \cite{twoIR}. 
More on the displacement pattern of the different modes is given in section IVD where the Raman spectra of the
three phases are compared.

Phonon dispersion relations along the high symmetry directions of the Irreducible Brillouin Zone (IBZ)
and phonon Density Of States (DOS) are reported in Fig. \ref{dos_alfa}.
The dynamical matrices have been computed for 30 {\bf q}  points along the $\Gamma-Z-R-X-\Gamma$ path shown in Fig. 
\ref{dos_alfa}, or on a $4\!\times\!4\!\times\!4$ grid in {\bf q}-space for the DOS calculation.
A Fourier interpolation technique provided the dynamical matrix at the other points of the BZ.\cite{dfpt}
For the DOS calculation,  we have used  
the tetrahedron method on a mesh of $20\!\times\!20\!\times\!20$ points in the IBZ.

An energy gap  separates  the low-frequency phonon branches  from the eight highest frequency bands, 
which can be assigned to stretching modes  of the eight inequivalent $\mathrm{Te-O}$ bonds per cell, i.e. 
symmetric and antisymmetric stretching modes of the four TeO$_2$ in the unit cell (see section IVD).
 This group of bands is slightly shifted
downward with respect to experiments, as discussed above.
Our results compare fairly well with previous lattice dynamics calculations 
of the DOS \cite{tesi,fononi_beta} and on dispersion relations. \cite{alfashell}
However, sizable quantitative differences are present especially on angular dispersion with respect to
shell model calculations. \cite{alfashell}
 To our knowledge, no experimental  data are available on phonon dispersion relations. 

\begin{table}
\caption{ Phonon frequencies (cm$^{-1}$) of $\alpha$-TeO$_2$
computed with BLYP functional at the experimental
lattice parameters (BLYPexp), with PBE iand LDA functionals at the theoretical lattice parameters
and with PBE at the experimental lattice parameters (PBEexp).}
\begin{ruledtabular}
 \begin{center}
  \begin{tabular}{l c c c c c c }
 Mode  & BLYPexp & PBEexp & PBE & LDA \\
   \hline
 $B_1$ (1) &   42 &   48 &   62 &   69 \\
 $A_2$ (1) &   90 &   94 &   90 &  106 \\
 $E$   (1) &  124 &  120 &  116 &  142 \\
 $B_1$ (2) &  128 &  127 &  125 &  129 \\
 $A_1$ (1) &  143 &  140 &  134 &  135 \\ 
 $B_2$ (1) &  148 &  141 &  136 &  131 \\
 $B_1$ (3) &  175 &  171 &  150 &  166 \\
 $E$   (2) &  175 &  171 &  162 &  179 \\
 $E$   (3) &  212 &  214 &  193 &  213 \\
 $A_1$ (2) &  214 &  207 &  195 &  206 \\
 $B_1$ (4) &  229 &  234 &  218 &  251 \\
 $A_2$ (2) &  269 &  267 &  254 &  256 \\
 $B_2$ (2) &  272 &  261 &  235 &  241 \\
 $E$   (4) &  291 &  282 &  262 &  262 \\
 $A_2$ (3) &  302 &  292 &  286 &  284 \\
 $E$   (5) &  325 &  316 &  295 &  327 \\
 $E$   (6) &  336 &  335 &  317 &  334 \\
 $A_1$ (3) &  383 &  377 &  364 &  402 \\
 $B_2$ (3) &  406 &  399 &  385 &  404 \\
 $A_2$ (4) &  535 &  507 &  515 &  471 \\
 $B_1$ (5) &  537 &  512 &  509 &  484 \\
 $E$   (7) &  594 &  573 &  567 &  551 \\
 $A_1$ (4) &  598 &  579 &  575 &  555 \\
 $E$   (8) &  700 &  683 &  694 &  667 \\
 $B_2$ (4) &  726 &  701 &  706 &  667 \\
  \end{tabular}
 \end{center}
\end{ruledtabular}
\label{confr_alfa}
\end{table}

\subsection{$\beta$-TeO$_2$}

Phonons at the $\Gamma$-point can be classified according to the irreducible representations of the
$D_{2h}$ point group  as $\Gamma=9(A_g+A_u+B_{1g}+B_{2g}+B_{3g)} + 8(B_{1u}+B_{2u}+B_{3u})$. The three translation modes
have been omitted.
The $g$-modes are Raman active while $B\ u$-modes are IR active. Only $B\ u$-modes couple to the macroscopic
longitudinal field via the non-analytic part of the dynamical matrix. 
The calculated phonon frequencies at the $\Gamma$-point, neglecting the contribution of the longitudinal macroscopic field
is given in Table \ref{phon_beta}.
The results refer to calculations at the experimental lattice parameters with the BLYP functional which turned out 
to provide the best agreement with experiments for $\alpha$-TeO$_2$ and for $\beta$-TeO$_2$ as well.

The Raman tensor for the active modes of an orthorhombic crystal has the form:\cite{loudon}

\begin{displaymath}
\begin{array}{rr}
A_g \Rightarrow
\left [ 
\begin{array}{ccc}
   a  &  ~.~ &  ~.~ \\
  ~.~ &   b  &  ~.~ \\
  ~.~ &  ~.~ &   c  \\
\end{array}
\right ]
\label{ag}
&
B_{1}(z) \Rightarrow
\left [ 
\begin{array}{ccc}
   ~.~  &  d &  ~.~ \\
 d  &   ~.~  &  ~.~ \\
  ~.~ &  ~.~ &   ~.~  \\
\end{array}
\right ]
\label{b1g}
\\ & \\
B_{2}(y) \Rightarrow
\left [ 
\begin{array}{ccc}
   ~.~  &  ~.~ &  e \\
 ~.~  &   ~.~  &  ~.~ \\
  e &  ~.~ &   ~.~  \\
\end{array}
\right ]
\label{b2g}
&\rule{5mm}{0pt}
B_{3}(x) \Rightarrow
\left [ 
\begin{array}{ccc}
   ~.~  &  ~.~ &  ~.~ \\
 ~.~  &   ~.~  &  f \\
 ~.~ &  f &   ~.~  \\
\end{array}
\right ]
\label{b3g}\\
\end{array}
\end{displaymath}
For $\beta$-TeO$_2$ only the $g$-modes have to be considered.

\begin{figure}
\caption{Unpolarized Raman spectra of polycrystalline $\alpha$-, $\beta$-, and $\gamma$-$\mathrm{TeO_2}$ for  backscattering geometry.
 Theoretical spectra are calculated with the BLYP functional at the experimental lattice parameters. 
 Experimental spectra  are also shown for comparison.\protect\cite{gamma2} For each phase the experimental
 spectrum is normalized over the strongest high-frequency peak of the theoretical spectrum.}
\includegraphics[width=1.0\columnwidth]{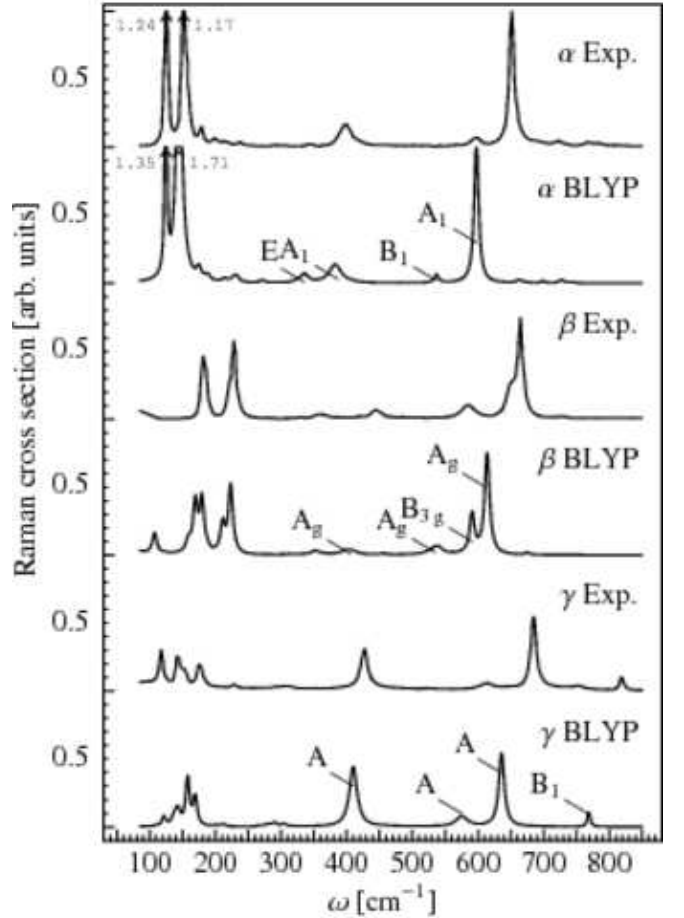}
\label{raman_blyp_exp}
\end{figure}

The calculated coefficients of the Raman tensor are given in Table \ref{phon_beta} for all active modes.
Experimental Raman spectra are available only for polycrystalline samples in backscattering geometry
for unpolarized light. \cite{gamma2} The corresponding theoretical spectrum can be obtained by averaging the Raman cross
section over the solid angle by making use of Eq. \ref{solidangle}. The resulting theoretical Raman spectrum of
$\beta$-TeO$_2$ is compared with the experimental powder spectrum in Fig. \ref{raman_blyp_exp}.
The linewidth of the most intense Raman peaks in the theoretical spectra has been obtained by fitting the experimental
spectra with a sum of Lorentzian functions. The results are given in Table \ref{phon_beta}.
The linewidth of the weaker or less identifiable peaks is set to 4 cm$^{-1}$. Overall the experimental spectrum is well reproduced, but for a redshift of the high frequency modes.
The underestimation of the frequency of the stretching modes is probably due to the overestimation of
the $\mathrm{Te-O}$ bond length (cfr. section IIIB). 
The frequency redshift with respect to experiments is even larger when the PBE functional is used, as occurs for the $\alpha$-$\mathrm{TeO_2}$ (cfr. Table \ref{confr_alfa}).
The theoretical Raman spectrum allows to assign the character of the modes responsible for the experimental peaks.
More on the displacement pattern of the different modes is given in section IV D where the Raman spectra of the
three phases are compared.

\begin{figure}
\caption{Angular dispersion of IR-active modes in $\beta$-$\mathrm{TeO_2}$.}
\includegraphics[width=1.0\columnwidth]{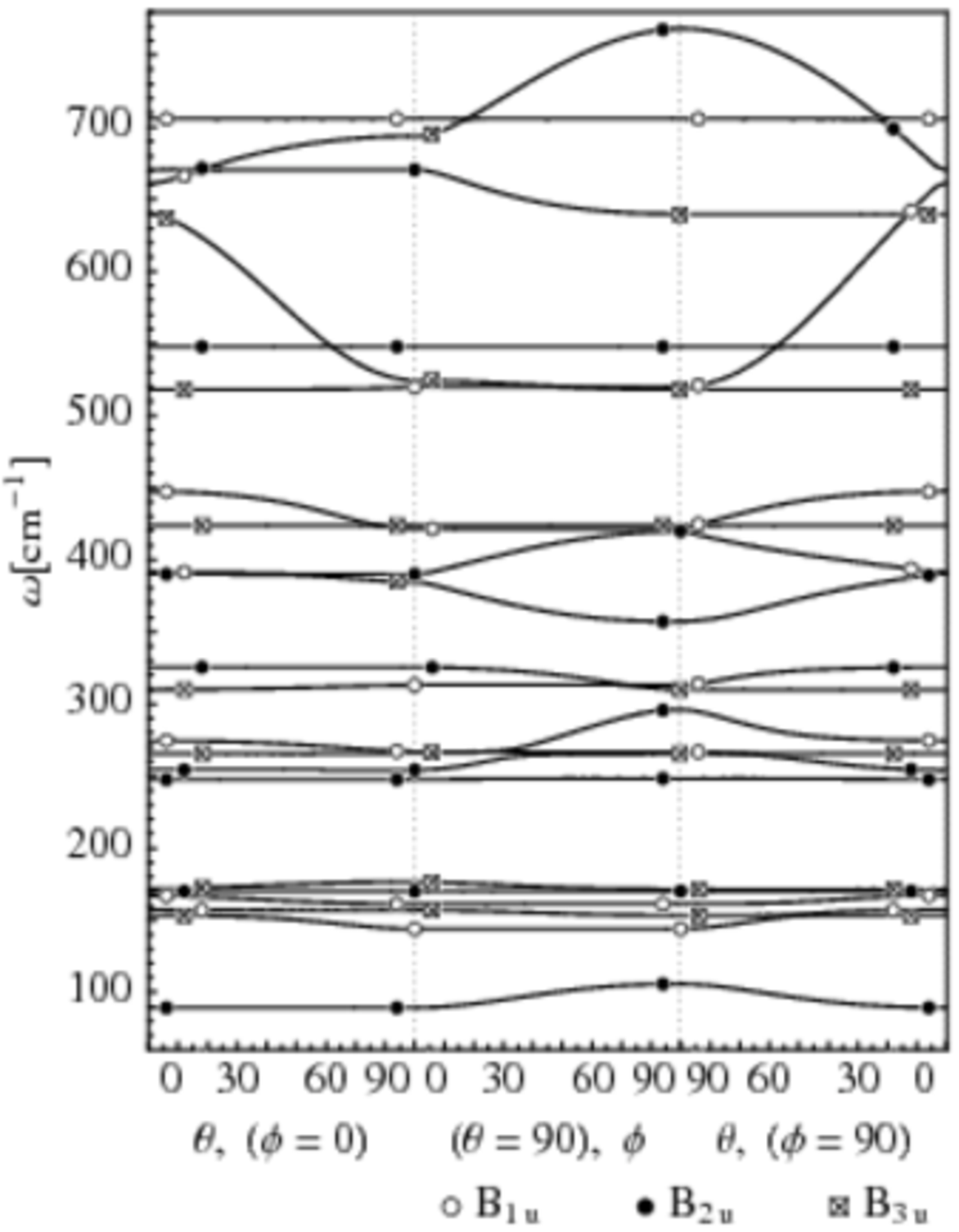}
\label{disp_beta}
\end{figure}

The $B\ u$-modes are subject to angular dispersion due to the coupling with the macroscopic longitudinal field.
The dependence of the phonon frequencies on the polar angles of the {\bf q} vector approaching the $\Gamma$-point is 
given in Fig. \ref{disp_beta}. 
 IR oscillator strength
of the active modes is reported in Table \ref{phon_beta}.
To our knowledge no experimental IR data are available  for single crystals. 
To compare with experimental data available we have thus  computed 
the IR absorption spectrum for a polycrystalline sample
as described in the previous section. 
The result is reported in Fig. \ref{poly_ir}; it has to be compared with the experimental spectrum reported in Fig. 2 of Ref. \cite{fononi_beta}.
The experimental peaks are reproduced qualitatively, but for an inversion in the relative intensities of the two
peaks around 600 and 700 cm$^{-1}$. Moreover, the theoretical spectrum shows a shoulder at high frequency which is absent in the
experimental spectrum. Our choice of the phonon linewidth might be partially responsible for these disagreements with the
experimental data. In fact,
in the lack of experimental data, 
the linewidth of the phononic modes is chosen similarly to those of $\alpha$-TeO$_2$, i.e. 5 cm$^{-1}$ for frequency below 150 cm$^{-1}$  and
15 cm$^{-1}$ for higher frequencies.

\begin{table*}
\label{phon_beta}
\caption[]{
 Theoretical phonon frequencies of $\beta$-TeO$_2$ at the $\Gamma$ point, oscillator strengths
 ($f_j$ in Eq. \protect\ref{absorb}) of IR active modes and squared coefficients of the Raman
 tensor of the Raman active modes (in units of $10^{-3}$\AA$^3$, see section IV).
 The contribution of the inner longitudinal macroscopic field is not included (LO-TO splitting).
 All the theoretical values correspond to calculations with the BLYP functional at the
 experimental lattice parameters.
 The phonon linewidths  ($\Delta\omega$) are obtained by fitting the experimental Raman spectra 
 of a polycrystalline sample \protect\cite{gamma2}  with a sum of Lorentzian functions 
(Fig. \protect\ref{raman_blyp_exp}).}
\begin{tabular*}{1.0\textwidth}{@{\extracolsep{\fill}}ccc}
\begin{tabular*}{0.46\textwidth}[t]{@{\extracolsep{\fill}}l c c c c c c}
\hline\hline
Mode &$\omega$ (cm$^{-1}$) & $f_j$ &  $a^2\ (d^2)$ & $b^2\ (e^2)$ & $c^2\ (f^2) $ &$\Delta\omega$\\ \hline 
$A_u(1)$      & 38         &            &           &          &           &      \\
$B_{3u}(1)$   & 44         & 0.049      &           &          &           &      \\
$B_{1u}(1)$   & 55         & 0.092      &           &          &           &      \\
$B_{3g}(1)$   & 56         &            &           &          &3.828      &      \\
$A_g(1)$      & 67         &            & 0.576     &1.607     &2.645      &      \\
$B_{1g}(1)$   & 72         &            & 0.000     &          &           &      \\
$B_{2g}(1)$   & 77         &            &           &0.310     &           &      \\
$A_u(2)$      & 85         &            &           &          &           &      \\
$B_{2u}(1)$   & 89         & 5.487      &           &          &           &      \\
$A_g(2)$      & 108        &            & 0.542     &3.133     &1.998      &      \\
$B_{2g}(2)$   & 123        &            &           &0.186     &           &      \\
$B_{1u}(2)$   & 144        & 5.503      &           &          &           &      \\
$B_{3u}(2)$   & 153        & 0.586      &           &          &           &      \\
$B_{1g}(2)$   & 159        &            & 0.612     &          &           &      \\
$B_{1u}(3)$   & 161        & 0.398      &           &          &           &      \\
$A_u(3)$      & 163        &            &           &          &           &      \\
$B_{3g}(2)$   & 169        &            &           &          &2.952      &  3.5 \\
$B_{2u}(2)$   & 170        & 0.015      &           &          &           &      \\
$B_{3u}(3)$   & 171        & 0.478      &           &          &           &      \\
$A_g(3)$      & 180        &            & 1.157     &2.548     &1.650      &  3.2 \\
$B_{2g}(3)$   & 188        &            &           &0.324     &           &      \\
$B_{3g}(3)$   & 208        &            &           &          &0.044      &      \\
$B_{1g}(3)$   & 209        &            & 0.003     &          &           &      \\
$A_g(4)$      & 211        &            & 1.163     &3.532     &2.730      &  3.4 \\
$B_{2g}(4)$   & 212        &            &           &0.257     &           &      \\
$B_{2g}(5)$   & 222        &            &           &0.049     &           &      \\
$A_g(5)$      & 223        &            & 7.267     &0.819     &1.616      &  4.4 \\
$B_{1g}(4)$   & 231        &            & 0.138     &          &           &      \\
$B_{3g}(4)$   & 234        &            &           &          &0.001      &      \\
$A_u(4)$      & 235        &            &           &          &           &      \\
$B_{2u}(3)$   & 247        & 0.666      &           &          &           &      \\
$B_{2u}(4)$   & 254        & 3.949      &           &          &           &      \\
$A_u(5)$      & 258        &            &           &          &           &      \\
$B_{3u}(4)$   & 265        & 0.094      &           &          &           &      \\
$B_{1u}(4)$   & 267        & 2.050      &           &          &           &      \\
\hline\hline
\end{tabular*} 
&  &
\begin{tabular*}{0.46\textwidth}[t]{@{\extracolsep{\fill}}l c c c c c c}
\hline\hline
Mode &$\omega$ (cm$^{-1}$) & $f_j$ &  $a^2\ (d^2)$ & $b^2\ (e^2)$ & $c^2\ (f^2) $ &$\Delta\omega$\\ \hline 
$B_{3u}(5)$   & 310        & 2.893      &           &          &           &      \\
$B_{1g}(5)$   & 312        &            & 0.171     &          &           &      \\
$B_{1u}(5)$   & 313        & 6.503      &           &          &           &      \\
$B_{3g}(5)$   & 317        &            &           &          &0.002      &      \\
$B_{2u}(5)$   & 325        & 0.940      &           &          &           &      \\
$A_u(6)$      & 336        &            &           &          &           &      \\
$B_{3g}(6)$   & 349        &            &           &          &0.000      &      \\
$B_{1g}(6)$   & 352        &            & 0.400     &          &           &      \\
$A_g(6)$      & 354        &            & 2.525     &0.078     &0.120      &      \\
$A_u(7)$      & 377        &            &           &          &           &      \\
$B_{2u}(6)$   & 390        & 0.455      &           &          &           &      \\
$B_{2g}(6)$   & 397        &            &           &0.127     &           &      \\
$A_g(7)$      & 404        &            & 0.112     &1.411     &4.031      & 13.4 \\
$B_{2g}(7)$   & 408        &            &           &0.129     &           &      \\
$B_{1u}(6)$   & 421        & 0.590      &           &          &           &      \\
$B_{3u}(6)$   & 424        & 0.000      &           &          &           &      \\
$B_{1g}(7)$   & 452        &            & 0.020     &          &           &      \\
$B_{3g}(7)$   & 457        &            &           &          &0.306      &      \\
$B_{3u}(7)$   & 518        & 0.115      &           &          &           &      \\
$B_{1u}(7)$   & 520        & 1.776      &           &          &           &      \\
$A_g(8)$      & 535        &            & 0.579     &1.192     &5.887      & 12.4 \\
$B_{2g}(8)$   & 541        &            &           &0.536     &           &      \\
$A_u(8)$      & 546        &            &           &          &           &      \\
$B_{2u}(7)$   & 548        & 0.000      &           &          &           &      \\
$B_{3g}(8)$   & 591        &            &           &          &8.965      &  8.7 \\
$B_{1g}(8)$   & 593        &            & 1.586     &          &           &      \\
$A_g(9)$      & 614        &            & 6.423     &18.540    &13.575     &  4.5 \\
$B_{2g}(9)$   & 615        &            &           &0.724     &           &      \\
$B_{3u}(8)$   & 640        & 0.618      &           &          &           &      \\
$A_u(9)$      & 668        &            &           &          &           &      \\
$B_{2u}(8)$   & 670        & 1.277      &           &          &           &      \\
$B_{3g}(9)$   & 675        &            &           &          &0.475      &      \\
$B_{1g}(9)$   & 675        &            & 0.023     &          &           &      \\
$B_{1u}(8)$   & 706        & 0.000      &           &          &           &      \\
\hline\hline
\end{tabular*}
\end{tabular*}
\end{table*}

\subsection{$\gamma$-TeO$_2$}

\begin{figure}
\caption{Angular dispersion of IR-active modes of $\gamma$-$\mathrm{TeO_2}$.}
\includegraphics[width=1.0\columnwidth]{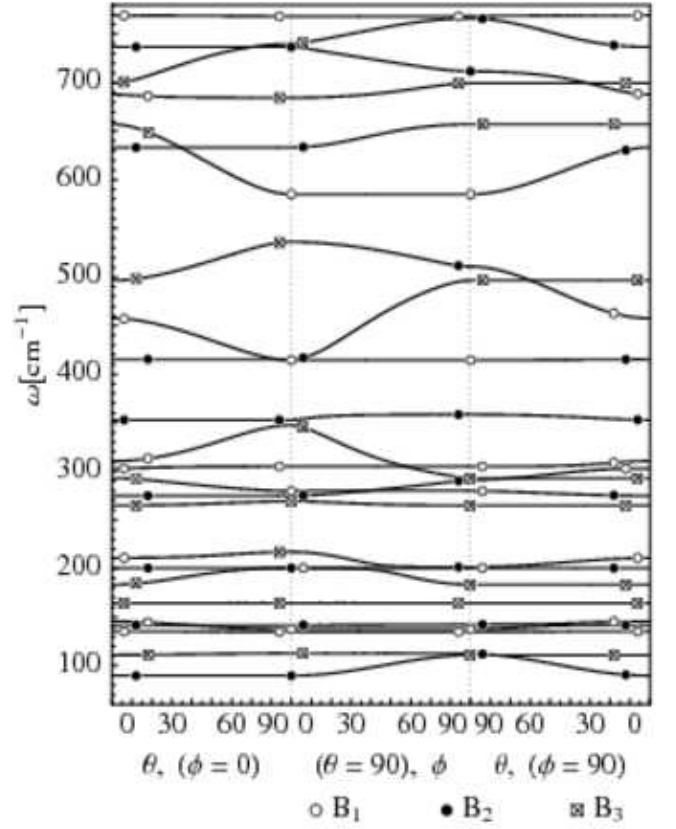}
\label{disp_gamma}
\end{figure}

Phonons at the $\Gamma$-point can be classified according to the irreducible representations of the
$D_2$ point group  as $\Gamma=8(B_{1}+B_{2}+B_{3)}) + 9 A$. The three translational  modes
have been omitted.
The calculated phonon frequencies at the $\Gamma$-point, neglecting the contribution of the longitudinal macroscopic field
is given in Table \ref{disp_gamma}.
The results refer to calculations at the experimental lattice parameters with the BLYP functional.
All the modes are Raman active while only $B$-modes  are IR active and display angular dispersion due to
the coupling with the longitudinal macroscopic field (Fig. \ref{disp_gamma}).
IR oscillator strength
of the active modes is reported in Table \ref{phon_gamma}.  Still
no experimental IR data are available  for $\gamma$-TeO$_2$  single crystals.
To compare with experimental data available we have thus  computed 
the IR absorption spectrum for a polycrystalline sample
as described in the previous section. 
The result is reported in  
 in Fig. \ref{poly_ir}; it has to be compared with the experimental spectrum reported in Fig. 2 of Ref. \cite{fononi_beta}.
In the lack of experimental data, 
the linewidth of the phononic modes is chosen similarly to 5 cm$^{-1}$ for frequency below 150 cm$^{-1}$  and
15 cm$^{-1}$ for higher frequencies, analogously to $\alpha$-TeO$_2$.
The theoretical spectrum reproduces all the main features of the experimental spectrum \cite{fononi_beta}, but for a redshift of
the theoretical modes at high frequency as occurs for $\alpha$-TeO$_2$.
The Raman tensor for active modes has been given in the previous section.

\begin{table}
\caption{
 Theoretical phonon frequencies of $\gamma$-$\mathrm{TeO_2}$ at the $\Gamma$ point, 
oscillator strengths
 ($f_j$ in Eq. \protect\ref{absorb}) of IR active modes and squared coefficients of the Raman
 tensor of the Raman active modes (in units of $10^{-3}$\AA$^3$, see section IV).
 The contribution of the inner longitudinal macroscopic field is not included (LO-TO splitting).
 All the theoretical values correspond to calculations with the BLYP functional at the
 experimental lattice parameters.
 The phonon linewidths  ($\Delta\omega$) are obtained by fitting the experimental Raman spectra   
 of a polycrystalline sample \protect\cite{gamma2}  with a sum of Lorentzian functions 
\label{phon_gamma}
(Fig. \protect\ref{raman_blyp_exp}).}
\begin{ruledtabular}
\begin{tabular}[c]{l c c c c c c }
Mode &$\omega$ (cm$^{-1}$) & $f_j$ &  $a^2\ (d^2)$ & $b^2\ (e^2)$ & $c^2\ (f^2) $ & $\Delta\omega$ \\\hline
$A(1)$     & 52         &            & 0.029     &0.813     &0.137      &      \\
$B_2(1)$   & 90         & 8.215      &           &0.002     &           &      \\
$B_3(1)$   & 111        & 0.549      &           &          &0.084      &      \\
$A(2)$     & 122        &            & 2.222     &1.502     &0.025      &  4.2 \\
$B_1(1)$   & 135        & 0.307      & 1.190     &          &           &  3.7 \\
$B_1(2)$   & 137        & 1.627      & 1.242     &          &           &  3.7 \\
$B_2(2)$   & 142        & 0.069      &           &0.076     &           &      \\
$A(3)$     & 158        &            & 0.693     &0.612     &2.512      &  4.5 \\
$B_3(2)$   & 164        & 0.144      &           &          &0.000      &      \\
$A(4)$     & 169        &            & 1.138     &2.259     &2.875      &  4.5 \\
$B_3(3)$   & 183        & 5.049      &           &          &0.003      &      \\
$B_2(3)$   & 200        & 0.141      &           &0.117     &           &      \\
$B_1(3)$   & 201        & 1.362      & 0.017     &          &           &      \\
$A(5)$     & 212        &            & 0.311     &0.455     &0.112      &      \\
$B_3(4)$   & 265        & 0.517      &           &          &0.264      &      \\
$B_2(4)$   & 275        & 1.935      &           &0.323     &           &      \\
$B_1(4)$   & 280        & 2.097      & 0.457     &          &           &      \\
$A(6)$     & 291        &            & 0.480     &0.529     &0.004      &      \\
$B_3(5)$   & 293        & 2.036      &           &          &0.661      &      \\
$B_1(5)$   & 305        & 0.037      & 0.387     &          &           &      \\
$B_2(5)$   & 353        & 0.608      &           &0.040     &           &      \\
$A(7)$     & 410        &            & 1.467     &0.794     &29.271     & 13.4 \\
$B_1(6)$   & 414        & 1.670      & 0.601     &          &           &      \\
$B_2(6)$   & 415        & 3.866      &           &0.257     &           &      \\
$B_3(6)$   & 496        & 0.895      &           &          &0.119      &      \\
$A(8)$     & 575        &            & 0.033     &0.151     &9.968      & 12.4 \\
$B_1(7)$   & 585        & 1.519      & 0.231     &          &           &      \\
$B_2(7)$   & 633        & 1.147      &           &0.006     &           &      \\
$A(9)$     & 636        &            & 10.494    &16.807    &3.977      &  8.7 \\
$B_3(7)$   & 657        & 0.577      &           &          &0.036      &      \\
$B_3(8)$   & 699        & 0.158      &           &          &0.026      &      \\
$B_2(8)$   & 737        & 0.088      &           &0.007     &           &      \\
$B_1(8)$   & 768        & 0.005      & 1.955     &          &           &  4.5 \\
\end{tabular}
\end{ruledtabular}
\end{table}

The coefficients of the theoretical Raman  tensor are given in Table XI for all the active modes.
As for $\beta$-TeO$_2$,  experimental Raman spectra are available only for polycrystalline sample in 
backscattering geometry. 
The corresponding theoretical spectrum can be obtained by averaging the Raman cross
section over the solid angle. However, as opposed to the case of  $\beta$-TeO$_2$, Raman active modes
in $\gamma$-TeO$_2$ can be subject to
angular dispersion (cfr. Fig. \ref{disp_gamma}
). As a consequence, formula Eq. \ref{solidangle} can not be used and the
integral over the solid angle has to be performed over  discrete angular values of the phononic wavevector {\bf q}
approaching the $\Gamma$ point. For each wavevector {\bf q} transfered in the scattering process the phonon eigenvector and frequency  
is computed by including the non-analytic part of the dynamical matrix.
The solid angle average is performed over 256 angles independent by symmetry. 
The calculated powder Raman spectrum of $\gamma$-TeO$_2$ is compared with the experimental spectra and with the spectra of the
other phases in Fig. \ref{raman_blyp_exp}. 
The linewidth of the theoretical peaks are fitted to experiments as discussed for $\beta$-TeO$_2$ in the previous section.
The linewidth of the weaker or less identifiable peaks is set to 4 cm$^{-1}$.
The linewidth of the Raman active modes are reported in Table XI.
The angular dispersion has a minor effect on the Raman spectrum since the most intense peaks turn out to
have $A$ character for which no coupling exists with the longitudinal macroscopic field:
the Raman powder spectra of $\gamma$-TeO$_2$ including or neglecting the angular dispersion are indistinguishable on the
scale of Fig. \ref{raman_blyp_exp}.
The agreement between theory and  experiments is overall good as for the other phases. Still a sizable  redshift
of the high frequency modes is present. 
Analogously to the other phases, the error is still slightly larger by making use of the PBE functional.
Moreover, a very weak experimental peak around 790 cm$^{-1}$ has no
counterpart in the calculated spectrum. Although three Raman active modes fall in that region, none of them has an
appreciable Raman intensity. 
More on the displacement pattern of the different modes is given in section IV D where the Raman spectra of the
three phases are compared.

\subsection{Comparison of Raman spectra of the three phases}

\begin{figure*}
\caption{(Color online) Displacement pattern of the phonons responsible for the main Raman peaks of 
polycrystalline $\alpha$-TeO$_2$ (cfr. Fig. \protect\ref{raman_blyp_exp}). The TeO$_2$ molecules formed by the two
shorter equatorial bonds are shadowed.}
\includegraphics[width=1.00\columnwidth]{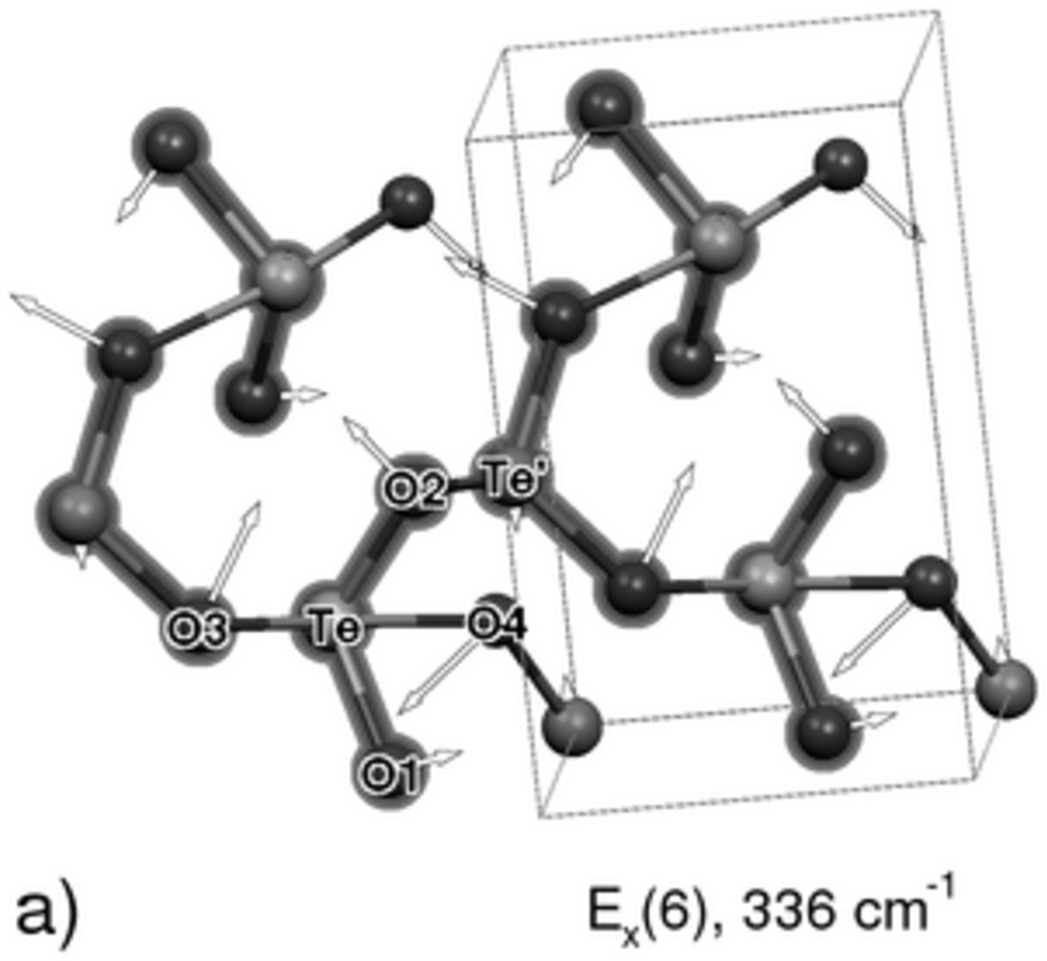}
\includegraphics[width=1.00\columnwidth]{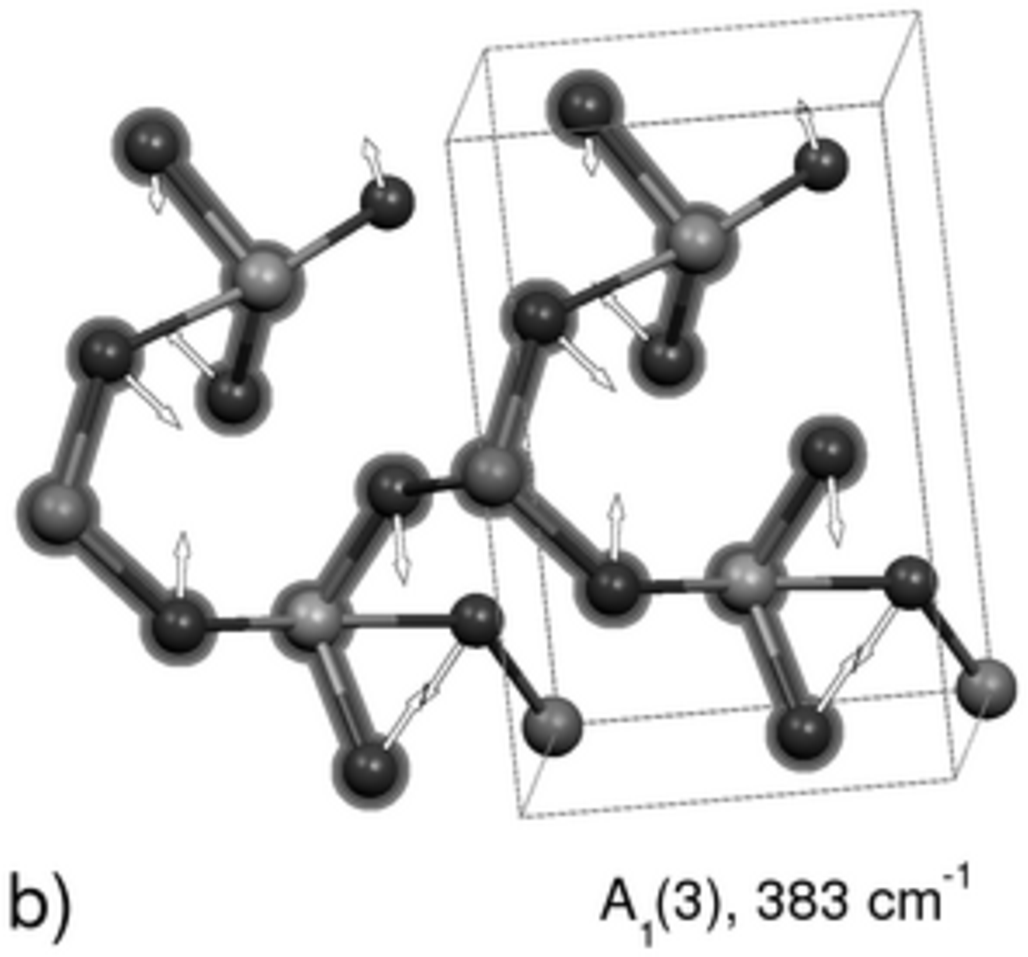}
\includegraphics[width=1.00\columnwidth]{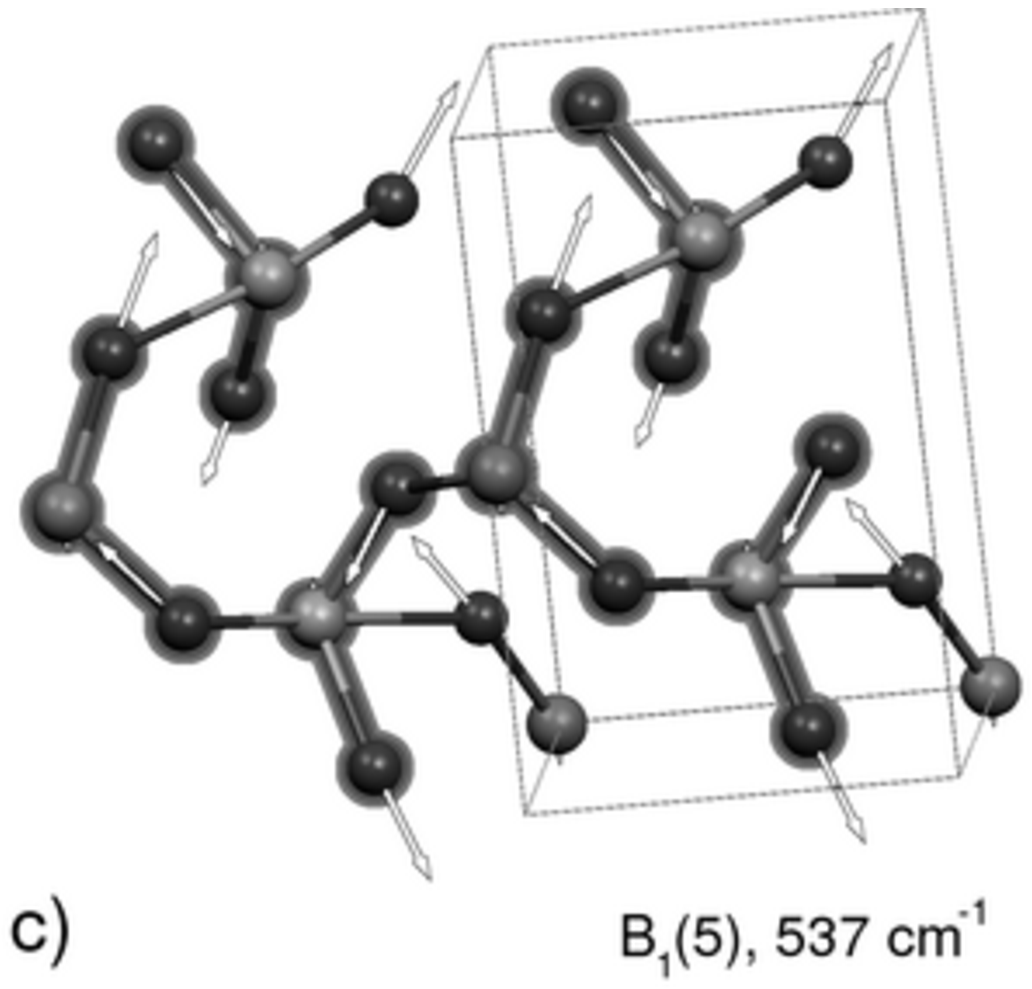}
\includegraphics[width=1.00\columnwidth]{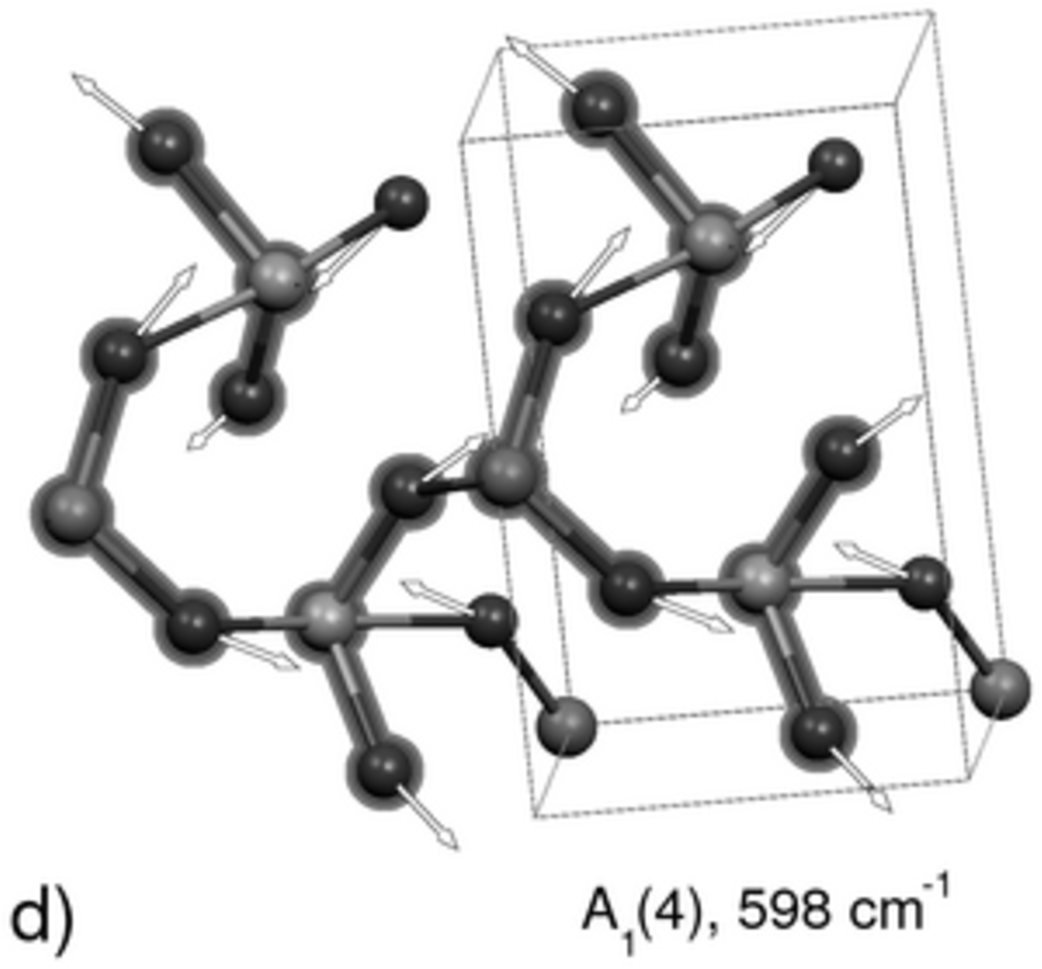}
\label{patterns_alfa}
\end{figure*}

\begin{figure*}
\caption{(Color online) Displacement pattern of the phonons responsible for the main Raman peaks of 
polycrystalline $\beta$-$\mathrm{TeO_2}$ (cfr. Fig. \protect\ref{raman_blyp_exp}).
The TeO$_2$ molecules formed by the two
shorter equatorial bonds are shadowed.}
\includegraphics[width=1.00\columnwidth]{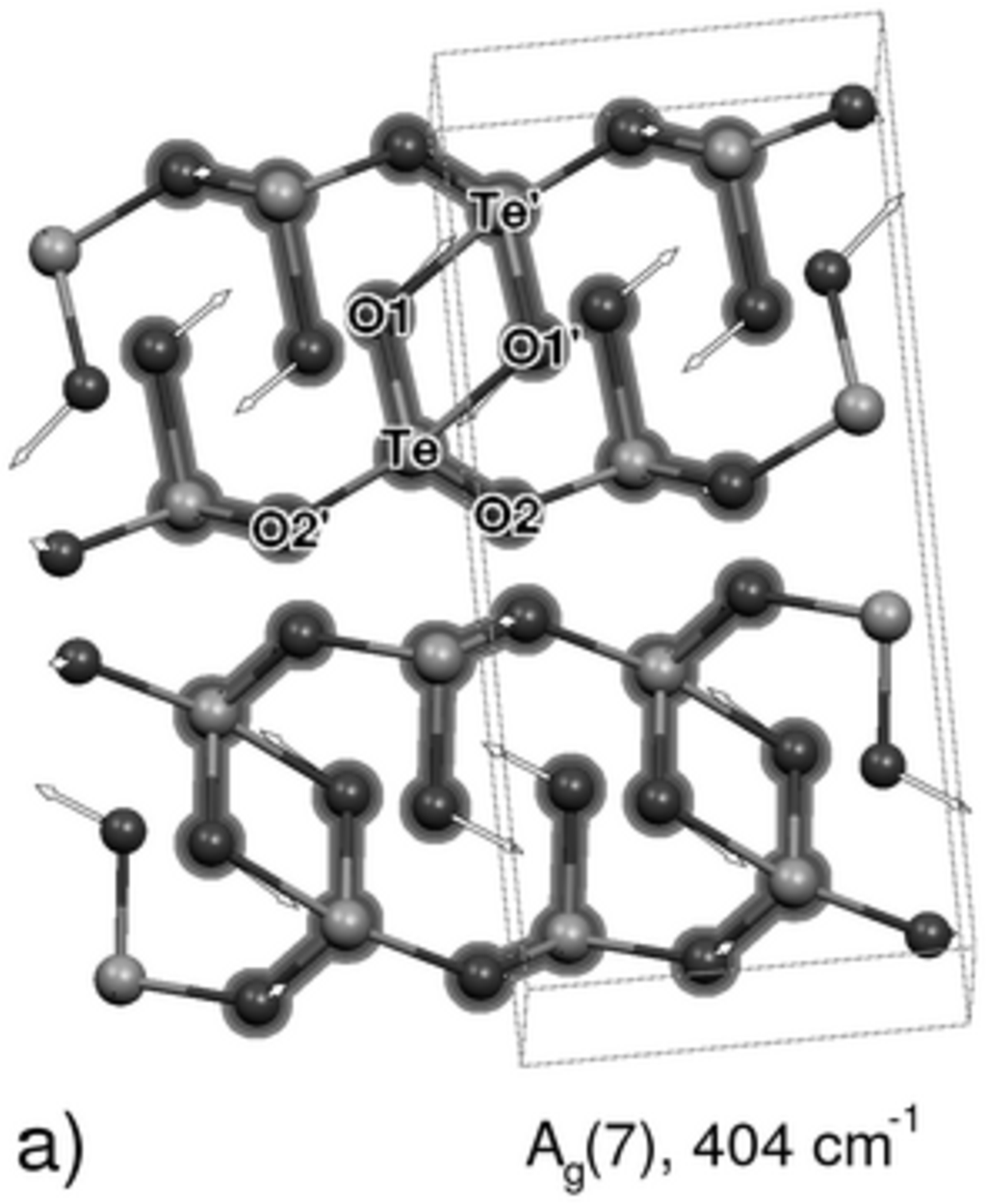}
\includegraphics[width=1.00\columnwidth]{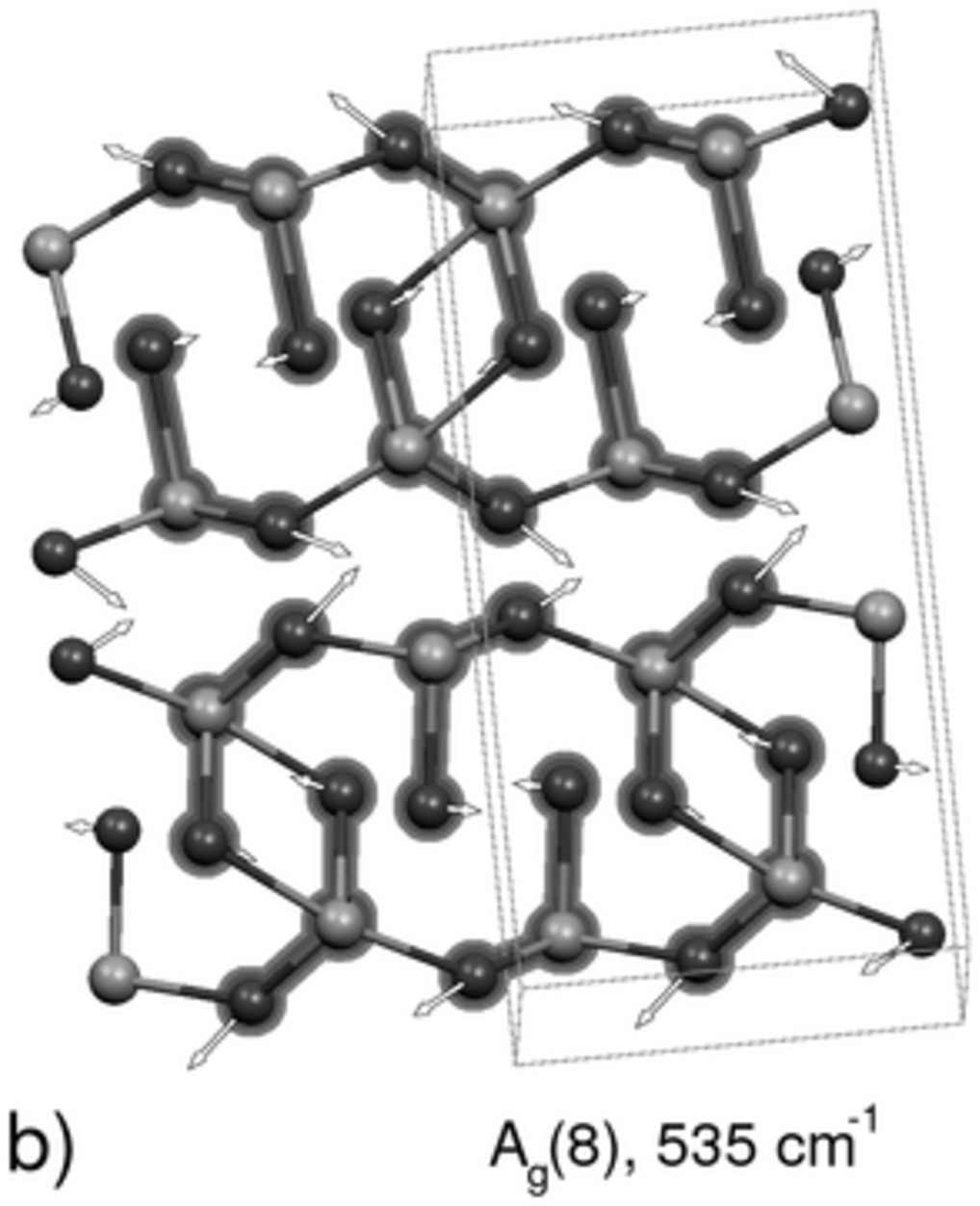}
\includegraphics[width=1.00\columnwidth]{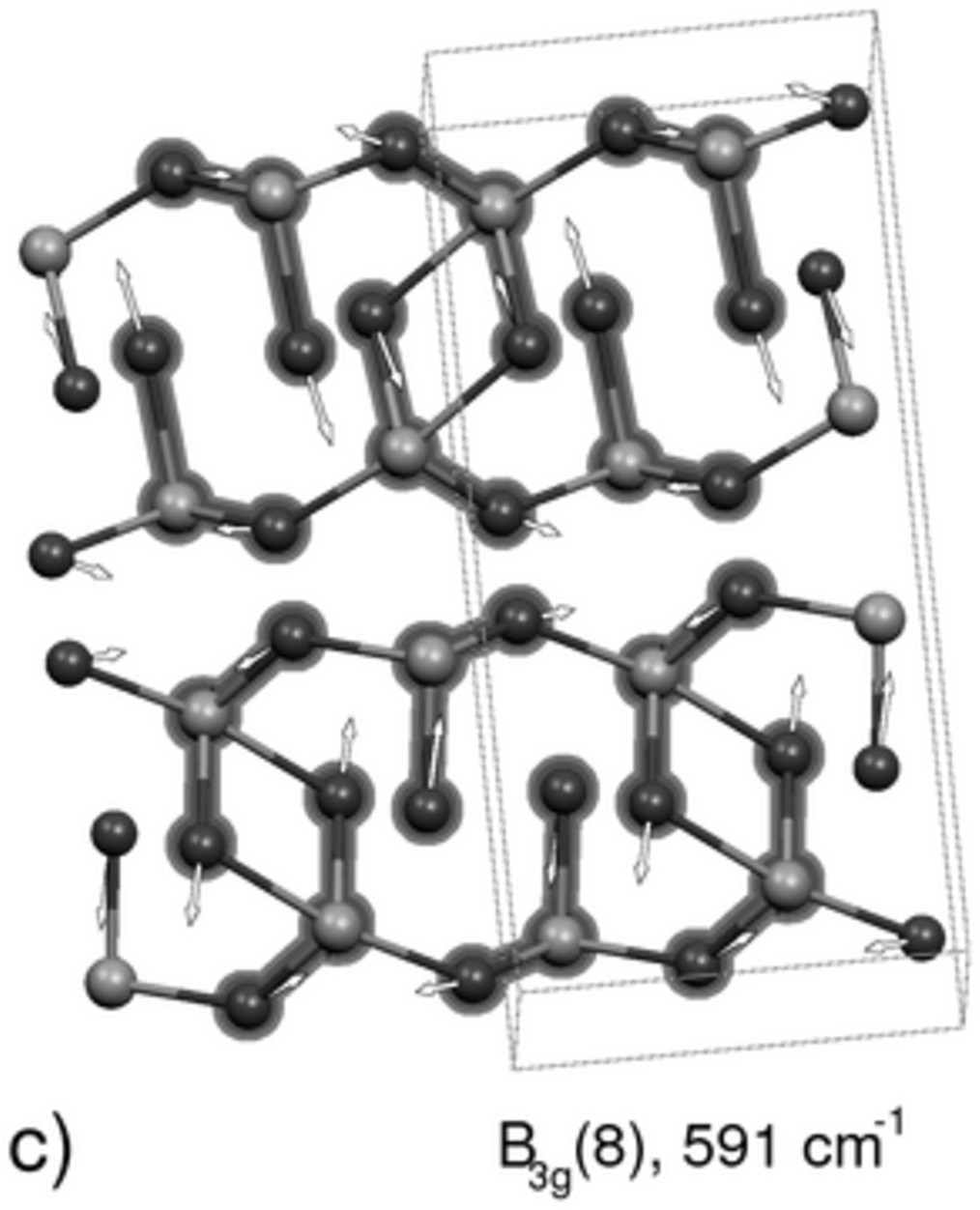}
\includegraphics[width=1.00\columnwidth]{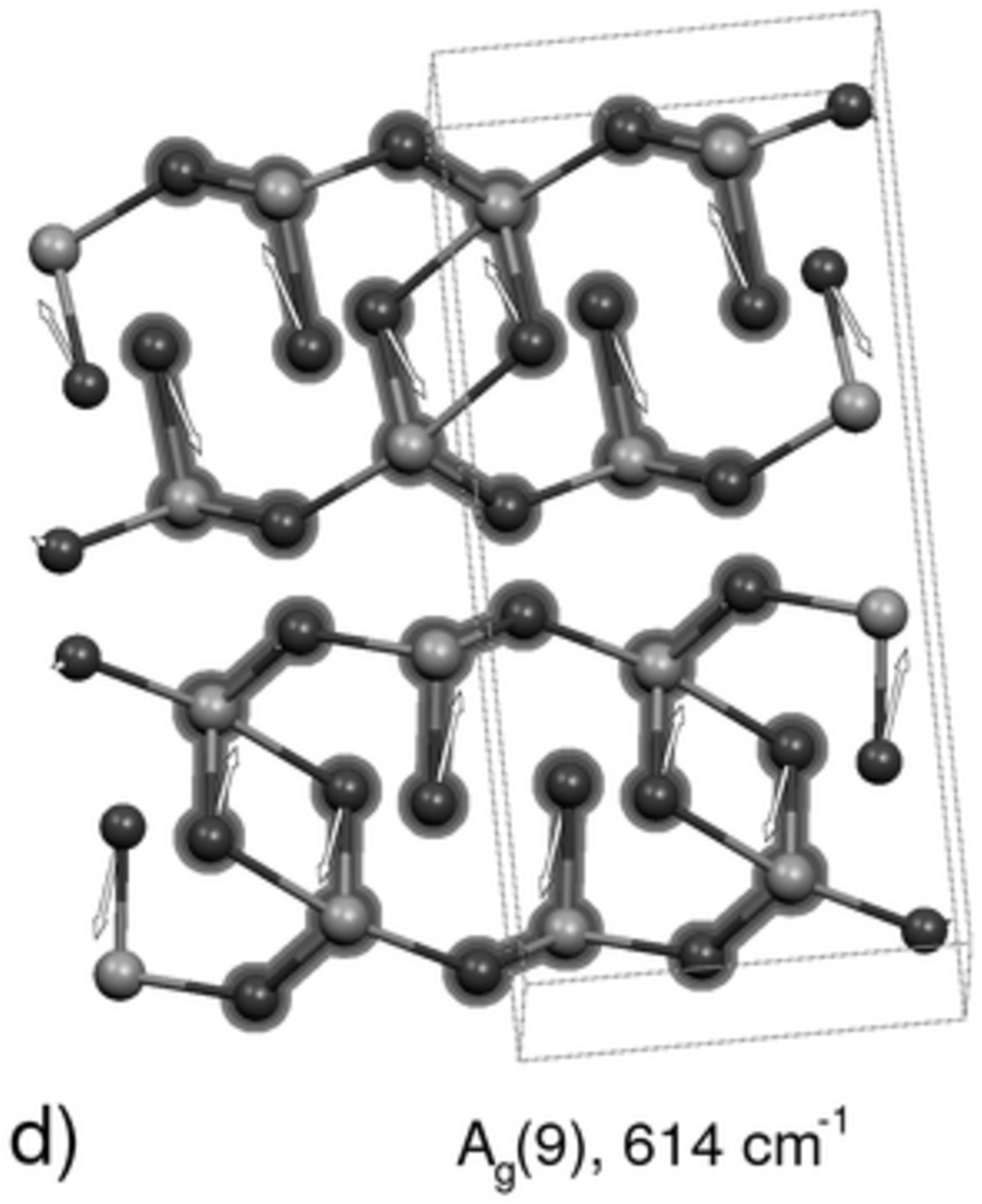}
\label{patterns_beta}
\end{figure*}

\begin{figure*}
\caption{(Color online) Displacement pattern of the phonons responsible for the main Raman peaks of 
polycrystalline $\gamma$-$\mathrm{TeO_2}$ (cfr. Fig. \protect\ref{raman_blyp_exp}).
The TeO$_2$ molecule formed by the two
shorter equatorial bonds are shadowed.}
\includegraphics[width=1.00\columnwidth]{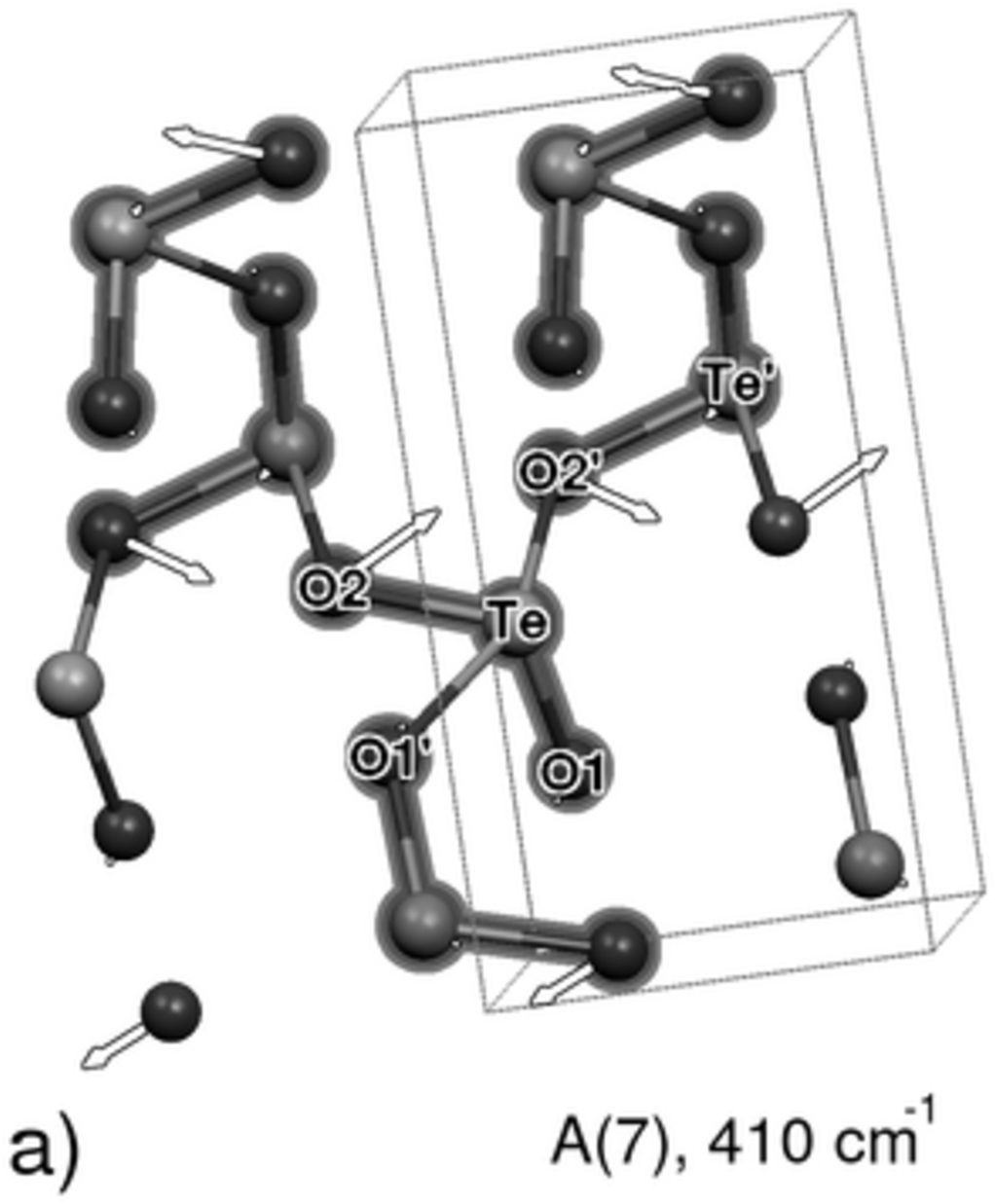}
\includegraphics[width=1.00\columnwidth]{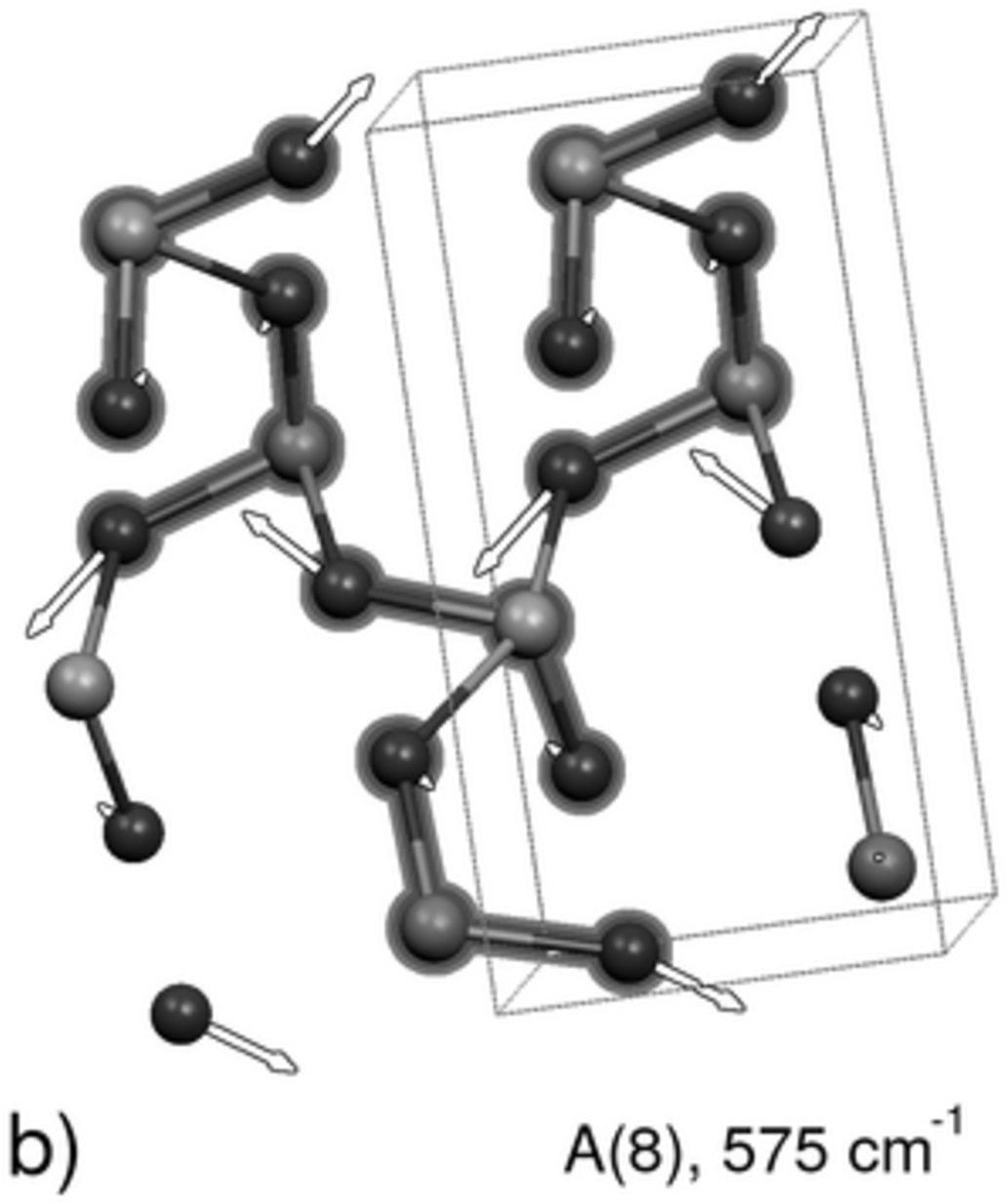}
\includegraphics[width=1.00\columnwidth]{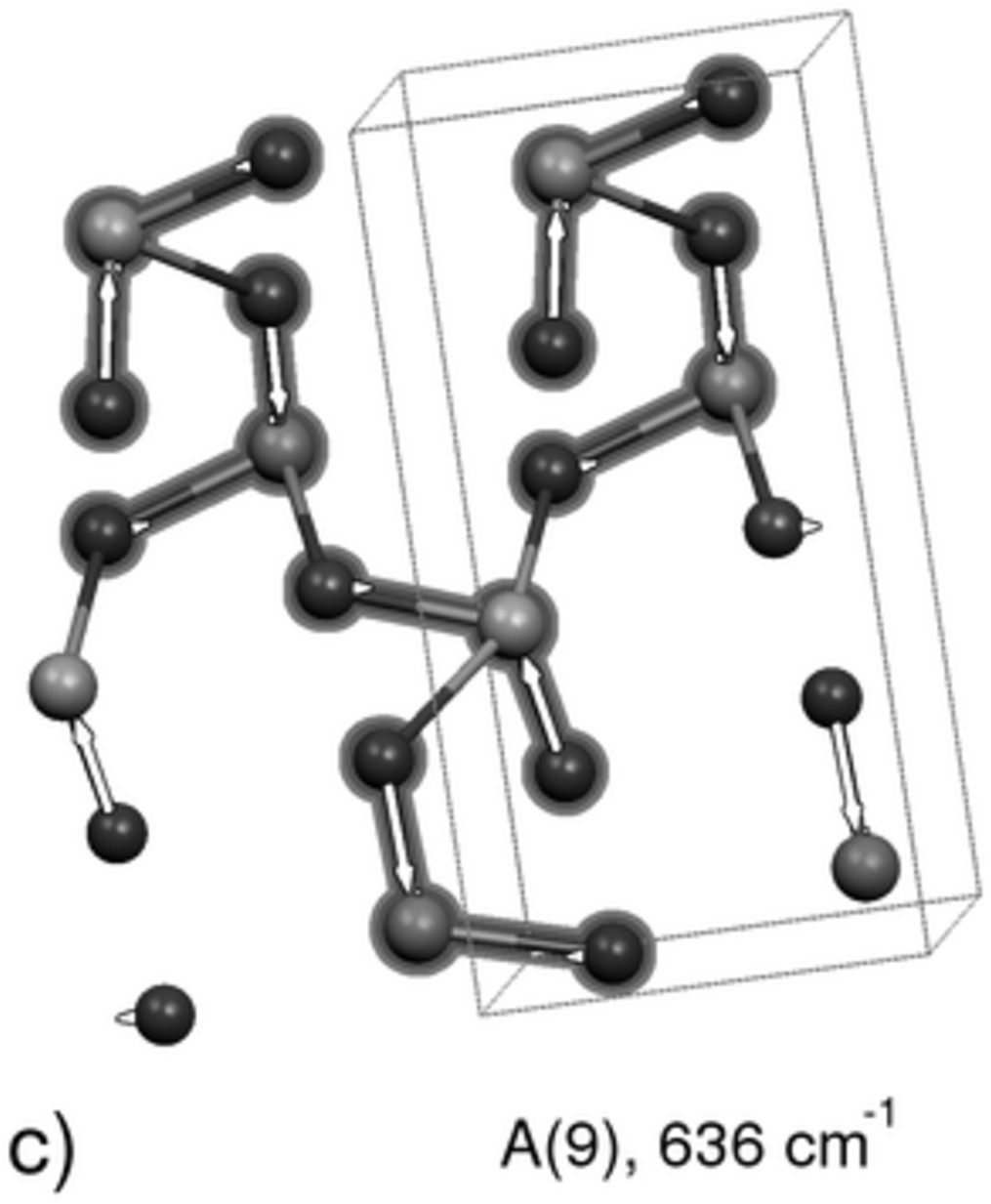}
\includegraphics[width=1.00\columnwidth]{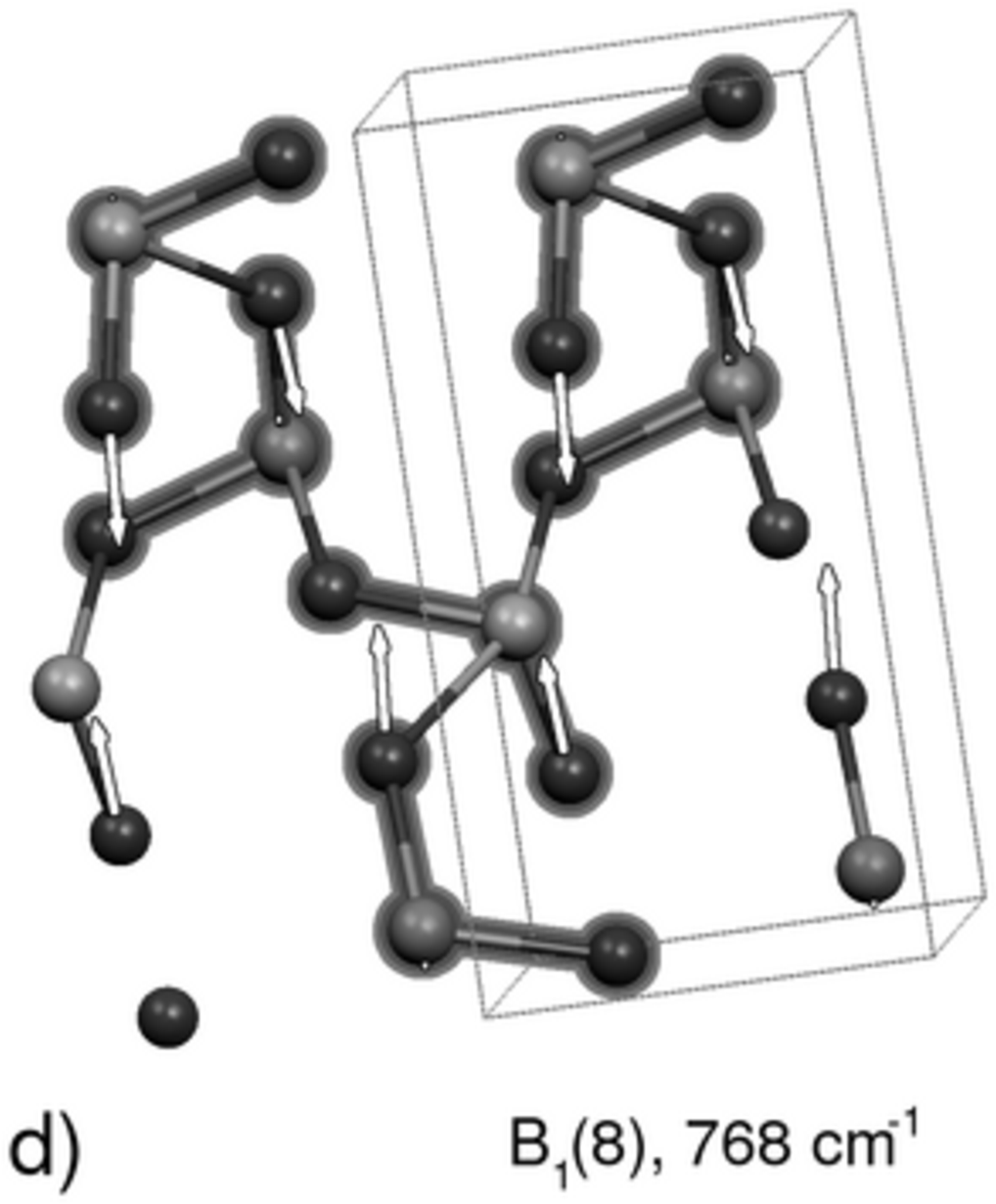}
\label{patterns_gamma}
\end{figure*}

In this section, we discuss the assignment of the experimental Raman peaks to specific phonons and
compare the displacements patterns of these phonons in the three different phases.
Firstly, we must say that the ab-initio results presented here are overall in good agreement with 
previous results from lattice dynamics calculations with empirical interatomic potentials.
\cite{gamma2,fononi_gamma,fononi_beta,tesi}
The displacement patterns of the phonons responsible for the main Raman peaks of
$\alpha$-, $\beta$-, and $\gamma$-TeO$_2$ are reported in Figs. \ref{patterns_alfa}, \ref{patterns_beta},
\ref{patterns_gamma}.

As pointed out in Ref. \cite{gamma2,tesi} the $\Gamma$-point phonons of $\alpha$-TeO$_2$ can be interpreted in terms
of the vibrations of the TeO$_2$  molecule given in Table I.
The modes $A_1$(4) and $B_1$(5) responsible for the strong Raman peak at 598 cm$^{-1}$ and the weak peak at
537 cm$^{-1}$ (cfr. Fig. \ref{raman_blyp_exp}) correspond to symmetric and antisymmetric stretching modes of the
TeO$_2$ molecular units (Fig. \ref{patterns_alfa}c-d).
The mode $A_1$(3) responsible for the Raman peak at 383 cm$^{-1}$ (Fig. \ref{raman_blyp_exp}) is a bending mode
of the TeO$_2$ molecule (Fig. \ref{patterns_alfa}b).
The mode $E$(6) responsible for the weak Raman peak at 336 cm$^{-1}$ is a librational mode of TeO$_2$ molecule which
modulates the length of the two axial bonds (Fig. \ref{patterns_alfa}a). 
The Raman peaks at lower frequencies (120-150 cm$^{-1}$) are due to librational modes of the TeO$_4$ units ($E$(1),
$A_1$(1), $B_2$(1) in Table VIII).
The displacement pattern shown in Fig. \ref{patterns_alfa} are in good agreement with those of the corresponding modes
obtained in Ref. \cite{gamma2,tesi} from lattice dynamics calculations with empirical interatomic potentials. 
However, the relative Raman intensities  are better reproduced by our ab-initio calculations than by the
Bond-Polarizability Model (BPM) used in Ref. \cite{gamma2,tesi}.

Concerning $\beta$-TeO$_2$, the mode $B_{3g}$(8) responsible for the Raman peak at 591 cm$^{-1}$ is an antisymmetric stretching
mode of the TeO$_2$ molecular unit (Fig. \ref{patterns_beta}c, in agreement with  Ref. \cite{gamma2,tesi}).
The mode $A_g$(9) responsible for the Raman peak at 614 cm$^{-1}$ is instead a stretching mode  of just one (the shorter) of the
two equatorial bonds (Fig. \ref{patterns_beta}d).
 The mode $A_g$(8) at 535 cm$^{-1}$ is the stretching mode of the other longer equatorial bond (Fig. \ref{patterns_beta}b).
This is in contrast with the results of lattice dynamics calculations with empirical potentials of Ref. \cite{gamma2,tesi},
which gives a mixing of the two modes to produce a symmetric and antisymmetric stretching mode of the TeO$_2$ molecule.
The mode $A_g$(7) at 404 cm$^{-1}$  is a vibration of the Te$_2$O$_2$ ring which modulates the length of the axial bonds,
in agreement with Ref. \cite{gamma2,tesi}. However, the bond-polarizability model used in Ref. \cite{gamma2,tesi} gives 
unsatisfactory results for the relative Raman intensities. 

Concerning $\gamma$-TeO$_2$, the Raman peaks at 768 and 636 cm$^{-1}$ correspond to stretching modes of the shortest $\mathrm{Te-O}$ bond 
in-phase ($A$(9)) or out-of-phase ($B_1$(8))
 within the four TeO$_2$ molecules in the unit cell, respectively (Fig. \ref{patterns_gamma}c-d).
The modes $A$(8) and $A$(7) responsible for the Raman peaks at 575 and 410 cm$^{-1}$ are mixed stretching and bending modes of the
$\mathrm{Te-O2'-Te'}$ bridge with the Te atoms fixed (Fig. \ref{patterns_gamma}a-b).
Again, the displacement  pattern of these modes are in good agreement with those calculated in Ref. \cite{gamma2,tesi} from
empirical interatomic potentials. 
The analysis of the vibrational spectra of $\alpha$-TeO$_2$ and $\beta$-TeO$_2$, presented above, identify clearly the
TeO$_2$  molecules as the building units of the crystal. However, the interaction between the molecules is large.
For instance, the symmetric stretching modes of the TeO$_2$ unit shifts from 598 cm$^{-1}$ ($A_1$(4)) to 726 cm$^{-1}$
($B_2$(4)) depending on the phase relation between the motion of the different molecules.
As pointed out in Ref. \cite{gamma2,tesi}, the $\gamma$-TeO$_2$ phase behaves differently with respect to 
$\alpha$- and $\beta$-TeO$_2$ for what concerns the vibrational properties.
Indeed, the displacement patterns reported in Fig. \ref{patterns_gamma} suggest a chain-like structure,
the most intense Raman peaks consisting of stretching modes of the $\mathrm{Te-O2'-Te'}$ bridge and of the shortest $\mathrm{Te-O}$ bond which
might be seen as side group of the polymeric chain. Still, the chains are strongly interacting since the
$A$(9) and $B_1$(8) modes which consists of the same intrachain vibration with different interchain phase relation differ
in frequency as much as 130 cm$^{-1}$. 
We note that in the $\beta$ phase the lowest energy phonons ($A_u$(1), $B_{3u}$(1), $B_{1u}$(1) at 38-55 cm$^{-1}$)
 correspond to rigid
out of-phase translational modes of the two layers in the unit cell.  
In  $\gamma$-TeO$_2$, the lowest mode  at 52 cm$^{-1}$ corresponds to a rigid translation of the chains along the
chain axes, out-of-phase between the two chains in the unit cell. The other two modes at higher frequency (80, 111 cm$^{-1}$)
involve instead sizable deformations of the chains themselves. 
In $\alpha$-TeO$_2$ the low frequency modes (below 140 cm$^{-1}$) can be well described as rigid motions of the TeO$_4$ units around
the bridging oxygen atoms.

\section{Conclusions}

Based on first principles calculations, we have studied the structural and vibrational properties of the three crystalline phases
of TeO$_2$ ($\alpha$, $\beta$, $\gamma$). Phonon dispersion relations and IR and Raman spectra have been computed within density
functional perturbation theory. The calculated Raman and IR spectra are in good agreement with available experimental data, 
and with previous lattice dynamics calculations based on empirical interatomic potentials \cite{gamma2,fononi_gamma,fononi_beta,tesi},
but for an
underestimation of the frequency of the Te-O stretching modes (above 500 cm$^{-1}$). The modes at high frequency (400-800 cm$^{-1}$)
can be well described in terms of vibrations of TeO$_2$ molecular units in $\alpha$- and $\beta$-TeO$_2$ which would then be seen as
molecular crystals. 
The analysis of the vibrational spectrum of $\gamma$-TeO$_2$ suggest instead a stronger intermolecular interaction supporting the
picture of $\gamma$-TeO$_2$ as a  chain-like structure, more connected than $\alpha$- and $\beta$-TeO$_2$.
However, the calculated elastic constants and electronic band structure do not show strong anisotropies which would have
 been expected from the description of $\gamma$-TeO$_2$ as the assembling of aligned polymers proposed experimentally. 
As a matter of fact, 
	TeO$_2$ molecules in $\alpha$- and $\beta$-TeO$_2$ and polymerized TeO$_2$ chains in $\gamma$-TeO$_2$ are strongly
interacting, possibly  via electrostatic coupling, as suggested   by
the strong ionic character of the Te-O bond emerged from previous ab-initio calculations.\cite{kleinman,palmero}

\section{Acknowledgments}

This work is partially supported by  the INFM Parallel Computing Initiative. Discussion with G. Dai and
F. Tassone are gratefully acknowledged.


\begin{thebibliography}{169}
\bibitem{o2te2} The calculated (PBE) bond lengths and vibrational frequencies of O$_2$ and Te$_2$ in the triplet states are respectively: O$_2$ d=1.254 \AA \ (exp. 1.208 \AA)\cite{handbook}, $\omega$=1570 cm$^{-1}$  (exp. 1580.2 cm$^{-1}$)\cite{handbook}; Te$_2$ d=2.584 \AA \ (exp. 2.557 \AA)\cite{handbook}, $\omega$=247 cm$^{-1}$  (exp. 247.1 cm$^{-1}$)\cite{handbook}.
\end{thebibliography}
\end{document}